\documentclass[conference]{IEEEtran}
\IEEEoverridecommandlockouts
\usepackage{cite}
\usepackage{amsmath,amssymb,amsfonts}
\usepackage{algorithmic}
\usepackage{graphicx}
\usepackage{textcomp}
\usepackage{xcolor}
\usepackage{subcaption}
\usepackage{xspace}
\usepackage{balance}

\def\BibTeX{{\rm B\kern-.05em{\sc i\kern-.025em b}\kern-.08em
    T\kern-.1667em\lower.7ex\hbox{E}\kern-.125emX}}

\newcommand{\voda}{VODA\xspace}
\newcommand{\vodaFullName}{Vulnerable Open-Source Dependency Analyzer\xspace}

\begin{document}

\title{A Comprehensive Study on the Impact of Vulnerable Dependencies on Open-Source Software\\
}

\author{    \IEEEauthorblockN{
    Shree Hari Bittugondanahalli Indra Kumar\IEEEauthorrefmark{1},
    Lília Rodrigues Sampaio\IEEEauthorrefmark{2},\\
    Andr{\'e} Martin\IEEEauthorrefmark{1},
    Andrey Brito\IEEEauthorrefmark{2},
    Christof Fetzer\IEEEauthorrefmark{1}
    }\\
    \IEEEauthorblockA{\IEEEauthorrefmark{1}Technische Universit{\"a}t Dresden, Dresden, Germany\\Email: shreehariuvce@gmail.com / andre.martin@tu-dresden.de / christof.fetzer@tu-dresden.de}
    \IEEEauthorblockA{\IEEEauthorrefmark{2}Universidade Federal de Campina Grande, Campina Grande, Brazil\\
    Email: liliars@lsd.ufcg.edu.br / andrey@computacao.ufcg.edu.br}\\
}

\maketitle

\begin{abstract}
Open-source libraries are widely used by software developers to speed up the development of products, however, they can introduce security vulnerabilities, leading to incidents like Log4Shell. With the expanding usage of open-source libraries, it becomes even more imperative to comprehend and address these dependency vulnerabilities. The use of Software Composition Analysis (SCA) tools does greatly help here as they provide a deep insight on what dependencies are used in a project, enhancing the security and integrity in the software supply chain.
In order to learn how wide spread vulnerabilities are and how quickly they are being fixed, we conducted a study on over 1k open-source software projects with about 50k releases comprising several languages such as Java, Python, Rust, Go, Ruby, PHP, and JavaScript. Our objective is to investigate the severity, persistence, and distribution of these vulnerabilities, as well as their correlation with project metrics such as team and contributors size, activity and release cycles. In order to perform such analysis, we crawled over 1k projects from github including their version history ranging from 2013 to 2023 using \voda, our SCA tool. Using our approach, we can provide information such as library versions, dependency depth, and known vulnerabilities, and how they evolved over the software development cycle. Being larger and more diverse than datasets used in earlier works and studies, ours provides better insights and generalizability of the gained results. The data collected answers several research questions about the dependency depth and the average time a vulnerability persists. Among other findings, we observed that for most programming languages, vulnerable dependencies are transitive, and a \textit{critical} vulnerability persists in average for over a year before being fixed. The results furthermore emphasize the importance of managing dependencies, performing timely updates, and suggests types of vulnerabilities that can be fixed faster.
\end{abstract}

\begin{IEEEkeywords}
security, vulnerable dependencies, software composition analysis
\end{IEEEkeywords}

\section{Introduction} \label{sec:intro}

Software development has considerably changed over time, going from code being written from scratch to rather a combination of custom code and open-source components. Such components are available to any developers, to either use or contribute to, in the form of libraries or packages. In modern software, we see that these components make up an average of $77\%$ of the code~\cite{synopsys2024}, increasing developers productivity by saving time and effort when reusing them. In order to do that, developers typically search for libraries and packages that meet their requirements and integrate them into their own code as dependencies. This process of automating the download and management of dependencies is signiﬁcantly simpliﬁed by package managers, available for a variety of programming languages, such as \textit{npm} for JavaScript, \textit{pip} for Python, \textit{maven} for Java, \textit{cargo} for Rust, and so on.

Although beneficial in many scenarios, the use of such open-source dependencies is one among numerous factors that have contributed to the growing complexity of modern software. For instance, developers must keep track of direct and indirect dependencies, their versions, and continuously update them. In this context, direct dependencies are the speciﬁc packages or libraries that a software directly depends on and are included in its own code, while indirect or transitive dependencies are the packages or libraries that are required by a software’s direct dependency but are not required directly by the software itself. Considering this, the managing of such libraries can be very complex, depending on how many are required by the software, and specially how many are actually transitive dependencies, which can easily be forgotten when simply manually tracking them. Besides that, other factors that increase software complexity are, ($1$)~frequent releases with shorter release development cycles, which can result in a rapidly changing software landscape, complicating the process of maintaining compatibility between releases; and ($2$)~a large number of contributors, which can make it diﬃcult to adhere to common guidelines and follow best software development practices. 

With increased complexity, there is also an increased probability that vulnerabilities are introduced into the software. This can happen in many different ways, for instance, if a vulnerability is unknown at the time of release, the software will be exposed to that until such vulnerability is discovered and the developers can mitigate it. In addition to that, if for some reason the latest version of (open-source) dependencies are not used by a particular software product, there is even a high probability that the software will contain a vulnerability as the support and vulnerability fixes are often only provided for the latest versions (or LTS versions, if available at all) of a package~\cite{Kula2018}. Another point to consider is that open-source components have often very complex inter-dependencies, and a software can be vulnerable regardless of vulnerabilities in its direct or transitive dependencies.

In general, software vulnerabilities can signiﬁcantly compromise the security of a system and make it susceptible to zero-day attacks. A zero-day attack takes advantage of a security ﬂaw that was previously unknown. Such vulnerabilities may allow attackers to gain unauthorized access to sensitive data, compromise systems, disrupt operations, or even seize control of a system for malicious purposes. To protect against zero-day attacks, it is essential to keep the system patched with the most recent security updates and to continuously monitor for new vulnerabilities, audit, and ﬁx them.

A recent example that caused a large impact on a huge number of software systems world wide was the vulnerability discovered in \textit{Apache Log4j}, an extremely popular logging library in the Java ecosystem. On December 2021, a new vulnerability (CVE-2021-44228~\cite{CVE-2021-44228}) was disclosed, which exposed the \textit{Log4j} package, disrupting the operation of systems all around the world. By injecting prepared strings into the logging library, an attacker could execute arbitrary code remotely. Remote Code Execution (RCE) can result in complete compromise of the target system, giving the attacker potential access to sensitive data and complete control of the system. RCE can have severe effects on the conﬁdentiality, availability, and integrity of the compromised system. Thus, according to the National Institute of Standards and Technology (NIST), the severity of the vulnerability was categorized as critical.

Google also conducted a vulnerability assessment~\cite{GoogleLog4j} and estimated that $17k$ Maven Central packages were affected. Millions of attempts to exploit corporate networks were made, with approximately $44\%$ of corporate networks~\cite{CheckpointLog4j} encountering such attempts. Consequently, it is evident that the impact of such a vulnerability can be catastrophic, and it is crucial that we are able to identify vulnerable components and resolve the security issue as soon as a new vulnerability is disclosed. Software Composition Analysis (SCA) tools can scan software and its dependencies, continuously check against a database of known vulnerabilities, such as the National Vulnerability Database (NVD)~\cite{NVD-2013}, and generate a report with potential security or compliance risks. Thus, incorporating SCA tools in the software development workflow increases the security and integrity levels of the software supply chain by ensuring third-party code does not introduce security risks into the system. 

Most SCA tools can generate a Software Bill of Materials (SBOM). An SBOM can be described as a detailed inventory of all dependencies and components referenced and used by an application. An ideal SBOM provides, for instance, the component’s name, date of release, checksum among other metadata for every component present in the evaluated system. This information is provided in different machine-readable formats, such as JSON and XML, following particular standards, the two primary ones being the Software Package Data Exchange (SPDX), promoted by the Linux Foundation, and CycloneDX, which originated from the Open Web Application Security Project (OWASP). Now, in the event of the Log4Shell vulnerability disclosure, nearly every organization affected by it immediately needed to discover which of its applications contained the compromised version of the Log4j library in order to mitigate its risk. Organizations with SBOMs identified the vulnerable component in a few hours, contrary to taking weeks to determine their risk~\cite{Roberts2021,Vaas2022}. Thus, the Log4Shell vulnerability showed that SBOMs can be useful for keeping track of and mapping applications that depend on vulnerable dependencies, and for shortening the time it takes to respond to the discovery of critical vulnerabilities.

Considering this, the goal of this work is to implement and evaluate a custom SCA tool, named \vodaFullName (\voda), which supports a wide variety of programming languages, including Java, Python, Go, Rust, Ruby, PHP, and JavaScript, and is later evaluated by running it against over one thousand popular open-source GitHub projects. Being larger and more diverse than datasets of earlier works on vulnerability dependency usage~\cite{Cadariu2015,Lauinger2018,Pashchenko2018,Prana2021}, our dataset enables better generalizability of the acquired results. To that end, we propose an event-driven, micro-service based architecture, composed of multiple micro-services loosely coupled via a RabbitMQ message broker. We also examine current SCA tools such as OWASP Dependency Track\footnote{OWASP, https://owasp.org/www-project-dependency-track/} and GitHub Dependabot\footnote{GitHub Dependabot, https://github.com/dependabot}, highlighting in which ways our tool fills the gaps they leave open. Besides that, as to facilitate application scaling and reproducibility of the experiments, the proposed \vodaFullName was deployed on a Kubernetes cluster, showcasing its distributed system nature.

Some of the key challenges in this process include: ($1$)~implementation of a highly scalable and distributed GitHub Crawler; ($2$)~generation of SBOMs for each project and its respective dependencies; and ($3$) analysis of the generated SBOMs. For ($1$), the GitHub Crawler considers all of the releases that are related to the open-source projects, which requires a high-level of data processing, and differs from approaches that only analyse the latest releases available. To acquire such data from GitHub, public REST endpoints are used, but in order to not to swamp the system with the distributed crawl of our system, a careful load balancing was implemented. There is also the need to define a metric for a repository’s measure of popularity, such as the number of stargazers and the number of contributors, from which we select whether to use one or a combination of the aforementioned attributes. As for ($2$), we use OWASP CycloneDX capabilities to facilitate the creation of an SBOM, however a few things still must be taken into account such as combining several SBOM files in case the CycloneDX plugin creates numerous SBOMs for a single repository. Regarding ($3$), to analyse such SBOMs, Sonatype OSS Index~\footnote{Sonatype OSS, https://ossindex.sonatype.org/} can be used to identify vulnerabilities, although it is a public service with rate limits on its REST APIs.

With the help of the data collected, we aim to answer a variety of research questions such as \textit{What is the current release trend in open-source applications and what are the most used dependencies in recent years?} \textit{What are the most prevalent vulnerabilities in open-source software, and what is the average time they persist in a system?} \textit{Are vulnerabilities really unknown at release time? How to prevent that?} In general, we consider repositories with releases made between 2013 and 2023, which have continuously grown over time, as results show. We also observed that Go presented more direct dependencies compared to Java, PHP, JavaScript, and Python, which have more transitive dependencies. 
We may claim that developers in the open-source ecosystem are not actively upgrading components to the most recent versions and, as a consequence, a vulnerability persists for an extended period of time.

The result of such an investigation can prove advantageous to both open-source library developers and users. By comprehending the frequency of persistent vulnerabilities and outdated dependencies, developers can encourage and facilitate updates. Users can also benefit from recognizing common vulnerabilities, which can help in anticipating and preventing them. This is particularly crucial since library vulnerabilities may not be immediately disclosed, and there may be a delay before developers provide a solution. Furthermore, the investigation's results can assist researchers in identifying research directions that are most likely to benefit a broad range of software projects.

The rest of the paper is organized as follows: Section~\ref{sec:background} introduces some fundamental concepts on dependency and vulnerability management, essential to the understanding of the solution. Section~\ref{sec:design} describes the architecture design of our \vodaFullName, highlighting its main components and SBOMs generation process, while Section~\ref{sec:implementation} details its implementation. Section~\ref{sec:evaluation} reports the results from the data collected after evaluating over $1k$ GitHub popular open-source projects with roughly $50k$ releases and Sections~\ref{sec:related}~and~\ref{sec:conclusions} enumerate related works and our conclusions, suggesting some future improvements.
\section{Background} \label{sec:background}

\subsection{SCA tools and SBOMs} \label{subsec:background:sbom}

Obtaining insight into the open-source components and dependencies used in a given application, as well as understanding the manner in which they are employed, is known as software composition analysis (SCA). This automated process serves the purpose of evaluating the security of these components and identifying any potential risks or licensing conflicts they may pose. By incorporating SCA tools into the software development workflow effectively, one takes a significant stride towards enhancing the security and integrity of the software supply chain. This ensures that any borrowed code does not introduce security risks or legal compliance issues into the developed products.

Performing a manual review of a dozen components may seem like a straightforward task. However, in today's software development landscape, applications are constructed using numerous libraries. These libraries, in turn, may have additional dependencies. As a result, the complexity of this process can extend to multiple layers and a seemingly small application could incorporate thousands of transitive dependencies. This is precisely where SCA tools prove invaluable.

There are various open-source tools available, including OWASP Dependency Check, Bundler-audit, RetireJS3, and Github Dependabot, which can be used to examine open-source dependencies for publicly-known security vulnerabilities. Additionally, several vendors, namely Sonatype, Synopsys and Veracode, provide software composition analysis tools. Besides identifying open-source libraries and pinpointing vulnerabilities associated with them, these SCA tools can determine the associated licenses and other metrics related to a software. Therefore, by using such tools, development teams can effectively identify vulnerable dependencies, as well as potential issues like outdated dependencies and license concerns.

One example of such SCA tools is OWASP Dependency Track, a user-friendly open-source artefact that offers a wide range of advanced features, such as vulnerability detection, policy management, and exploit prediction, as well as notification support. It has seamless integration with various vulnerability databases, including NVD, GitHub Advisory, Sonatype OSS Index, VulnDB, and others. On the other hand, GitHub Dependabot is adaptable to any GitHub repository and provides relevant insights on vulnerable repository components. It relies on the GitHub Advisory Database~\cite{GithubAdvisoryDatabase} as its primary data source. Furthermore, users can configure it to automatically update vulnerable component versions to the latest version.

For our use case, however, to effectively analyze the evolution of vulnerable components in a large number of software products accessible through git repositories, we intend to evaluate every release, which results in tens of thousands checks which requires a scalable software solution. While OWASP Dependency Track is a suitable option for monitoring the main branch of a few repositories typically used in mid-sized enterprises, considering multiple releases may require creating multiple projects for each release, resulting in time-consuming processes and compromising the software's ease of use. Additionally, neither OWASP Dependency Track nor GitHub Dependabot offer a date-based filtering option, preventing us from filtering and identifying vulnerabilities known at the time of release. Instead, we are provided with a list of all vulnerabilities discovered to date, which prevents a detailed study on the evolution of these vulnerabilities over time, a key aspect of our research. To address these challenges, we have implemented \voda, our \vodaFullName, which provides features to facilitate the processing of thousands of releases in parallel and enables an easy analysis of past data.

It is important to note that most SCA tools can generate a software bill of materials (SBOM), which is a formal collection of machine-readable metadata that uniquely identifies a software package and its associated components. An apt comparison is to envision an SBOM as a comprehensive list of ingredients for a product, as it should encompass all the components upon which the software relies. These components may consist of third-party elements such as library modules or an application framework. 

At the bare minimum, an SBOM includes the component name, publisher name, component version, dependency relationship, author of SBOM data, file name, license information, and time stamp. The Open Web Application Security Project CycloneDX, the Software Product Data Exchange (SPDX), and the Software Identification Tagging (SWID) are three widely known SBOM standards, which is a schema designed to provide a common format for describing the composition of software in a manner that can be easily interpreted by machines and utilized by various other tools.

Our work leverages OWASP CycloneDX SBOMs, which is a lightweight SBOM standard designed to enhance supply chain capabilities for mitigating cyber risks. CycloneDX offers extensive tool support for creating an SBOM for various programming languages. Additionally, the OWASP CycloneDX BOM format is supported by several Software Composition Analysis (SCA) tools, including OWASP Dependency Track~\cite{OWASPDT}.

A standard OWASP CycloneDX SBOM can be represented in JSON, XML, or Protocol Buffers, encompassing various subsections such as BOM metadata, components, services, dependency relationships, compositions, vulnerabilities, and extensions. The \textit{BOM metadata} provides information about the source, manufacturer, tools utilized for BOM production, and licenses. \textit{Components} contain specific details regarding the software's dependencies, including coordinates (name, group, version), package URL, Common Platform Enumeration (CPE), SWID, and cryptographic hash. \textit{Services} refer to external APIs that the software may invoke. The \textit{dependencies} section delineates the direct and transitive relationships between components.

OWASP CycloneDX has a rich support when it comes to creating an SBOM, comprising specific plugins depending on the programming language of the given software. Available plugins include CycloneDX Maven, Gradle, Python, Go, Cargo, Ruby, Composer and Node NPM, which offer support to, respectively, Java, Python, Go, Rust, Ruby, PHP and JavaScript programming languages. These plugins usually generate SBOMs from files in the repository that indicate in some way the packages used by the software, such as \texttt{pom.xml}, \texttt{build.gradle}, \texttt{requirements.txt}, \texttt{go.mod} and so on.

However, the goal of SCA process and tools go beyond scanning application sources and binaries to produce an SBOM. The real challenge is accurately mapping each component version to known vulnerabilities, which can be done by reviewing CVE feeds from MITRE or NVD. SCA tools must be intelligent enough to map security vulnerabilities to impacted components and avoid flagging benign ones. To ensure developers are not burdened and compliance teams are at ease, SCA solutions must minimize false-positive vulnerabilities without introducing false negatives (i.e., missing security risks).

\subsection{Vulnerabilities databases}

Information on known software security vulnerabilities can be found in vulnerability databases. These databases are managed by organizations, security researchers, and software vendors to offer users a centralized resource for understanding the potential risks associated with using software that has known vulnerabilities, as well as identifying the specific versions or patches that address these vulnerabilities. A scoring system is utilized to categorize each vulnerability according to its severity.

The Open Source Vulnerability Database (OSVDB), National Vulnerability Database (NVD), X-Force, CERT-VN, and CISCO IntelliSchield Alerts are among the publicly accessible vulnerability databases. They all use the Common Vulnerability Scoring System (CVSS) as the preferred standard for rating the severity of a vulnerability. According to Johnson et al.~\cite{Johnson2018}, CVSS is a dependable and robust grading method for rating the severity of a vulnerability, with the NVD being ranked as the most effective vulnerability database.

The National Vulnerability Database (NVD) is maintained by the National Institute of Standards and Technology (NIST) and is widely recognized as a highly credible source of vulnerability information. It provides comprehensive details on various vulnerabilities, including those discovered and disclosed by researchers as well as those reported to software vendors. Each vulnerability listed in the NVD is assigned a unique identifier, severity rating, and other pertinent information. The NVD is regularly updated with the latest information and offers a plethora of data to assist both businesses and individuals in assessing security risks and making informed decisions.

The severity of vulnerabilities is assessed using the CVSS, which is widely recognized as the industry standard. The management of CVSS falls under the responsibility of the Forum of Incident Response and Security Teams (FIRST) group, an organization that specializes in incident response and security. Over time, CVSS has undergone multiple revisions, with the most recent iteration being \textit{v3.1}. The scores assigned by CVSS range from $0$ to $10$, and the severity levels can be categorized as \textit{Low, Medium, High,} or \textit{Critical}, depending on the score.

    
The scores, in their turn, can be categorized as follows:
\paragraph{Base score} The vulnerability's unchanging features across various user environments are represented by the base score. This score is determined through the use of base metrics, which consist of Exploitability and Impact metrics, and consider the effects on confidentiality, integrity, and availability. The base score, which ranges from $0$ to $10$, indicates the severity of the vulnerability, with higher numbers indicating greater severity.
\paragraph{Temporal score} The dynamic nature of vulnerability characteristics across various user environments and time is represented by the temporal score. This score factors in the current state of the threat landscape, the existence of patches, the presence of attack scripts, and the probability of exploitation. The score is also measured on a scale of $0$ to $10$.
\paragraph{Environmental score} Environmental metrics are applicable to the particular context where a vulnerability is present. These metrics are inherently unique to each individual enterprise. They pertain to either the level of importance the asset holds for the business or the measures taken to counteract or reduce the organization's vulnerability. The scores are measured on a scale of $0$ to $10$.

\section{Design of the \vodaFullName} \label{sec:design}

\voda (\vodaFullName) aims to facilitate the tracking of dependencies and its vulnerabilities across thousands of open-source repositories simultaneously, considering previous and current releases. In order to achieve that, we designed the system with the following components: ($1$)~a GitHub crawler capable of gathering information about each repository and its associated releases, using the number of stargazers as a measure of popularity; ($2$)~a programming language agnostic SBOM generator that internally uses the abundance of tools provided by CycloneDX; ($3$)~an SBOM analyser to identify vulnerabilities related to every component and produce a proper report with meaningful insights. 

In this section we describe these components and its adjacent services, an overview of the tools integrated into the system, and present explanations for the design decisions and reasoning behind them. 

\subsection{Key challenges and architecture details}

To implement \voda, there are a number of challenges to overcome: For instance, in order to evaluate the system, we want to use a very large number of notable open-source projects, built in many different programming languages (Java, Python, Go, Ruby, Rust, PHP, and JavaScript). We define as notable projects which are widely used such as log4j, a common logging facility in Java projects. Such projects are typically starred on platforms such as Github etc. We can acquire such data from source code repository hosting platforms such as GitHub, gitlab, bitbucket etc. by using public REST endpoints, but should be aware that these platforms are not keen to be swamped by the requests of our distributed system, hence, implement some request rate limiting. There is also the need to define a metric for a repository’s measure of popularity, such as the number of stargazers and the number of contributors in order to decide to include a repository in the crawl and subsequent analysis or not. 

\begin{figure}
    \centering
    \includegraphics[width=0.9\linewidth]{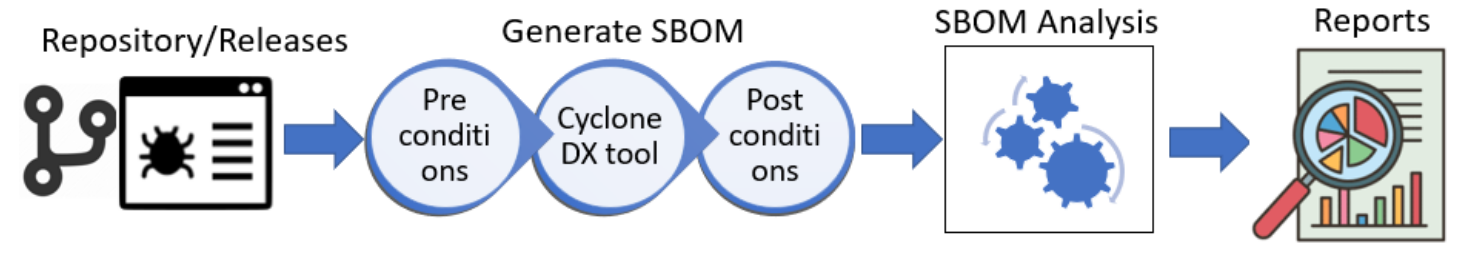}
    \caption{Key goals of \voda.}
    \label{fig:key-challenges}
\end{figure}

A simple overview of our system's workflow can be seen in Figure~\ref{fig:key-challenges}. In addition to the workflow, we face the challenge of the SBOM generation for multiple languages.
Fortunately, OWASP CycloneDX offers a multitude of tools to facilitate the creation of SBOMs for various languages, however, each tool has its own requirements that must be met in order to successfully produce an SBOM. A Go repository, for instance, may not have a \texttt{go.mod} ﬁle. In such situations, we must generate an appropriate ﬁle to meet the prerequisites. In addition, if the CycloneDX plugin creates numerous SBOMs for a single repository, these files must be combined in order to produce a single SBOM ﬁle. Moreover, the generated SBOM files must be inspected for the system's related dependencies and vulnerabilities in an eﬃcient manner, for instance by querying Sonatype OSS Index. However, in order to not to overload such public services, we require strategies such as rate limiting for the publicly available REST APIs interface, and caching in order to reduce the overall lookup frequency.

\begin{figure}
    \centering
    \includegraphics[width=0.8\linewidth]{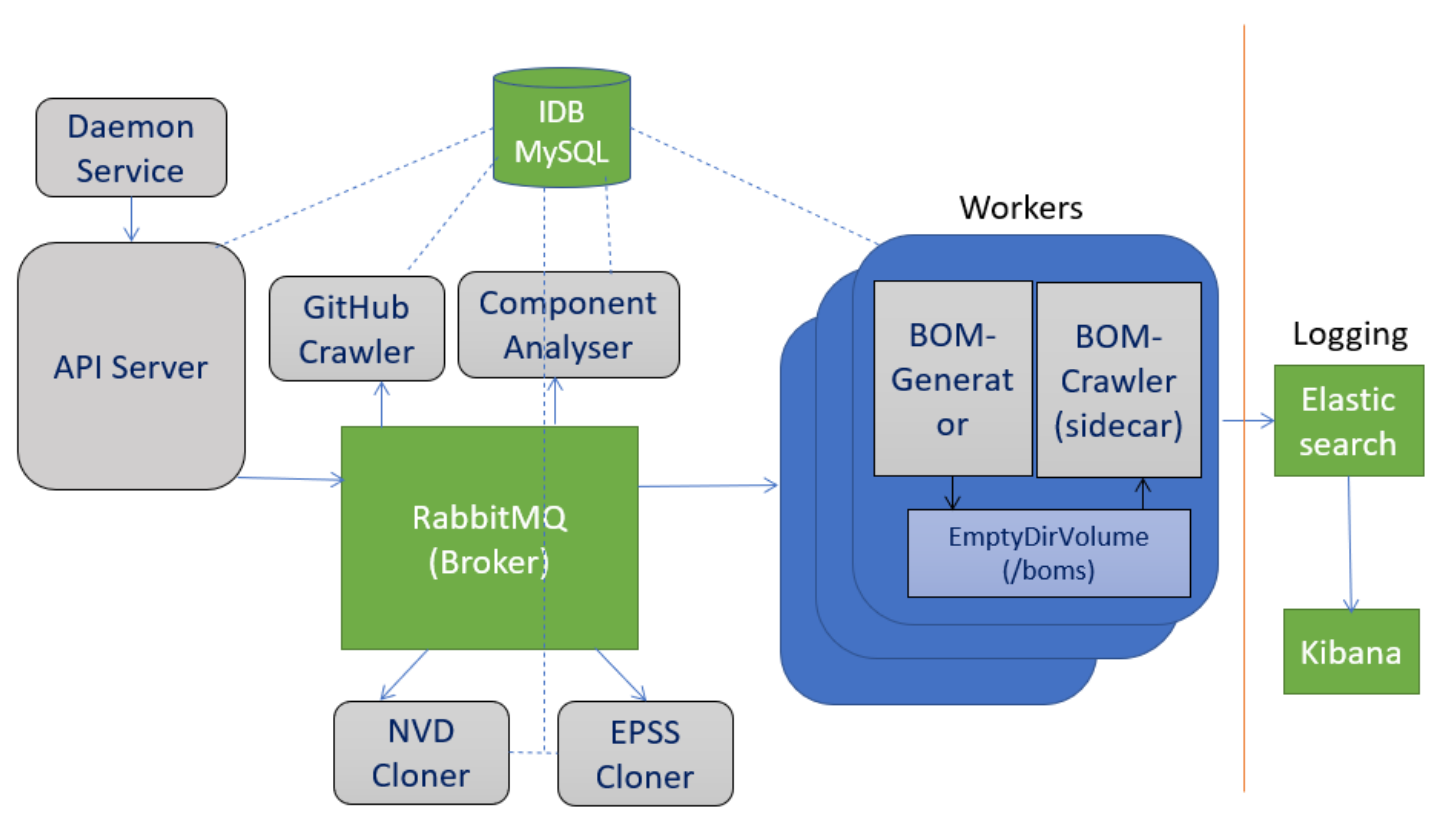}
    \caption{Overview of the \voda architecture.}
    \label{fig:arch-overview}
    \vspace{-1.5em}
\end{figure}

Considering this, as depicted in Figure~\ref{fig:arch-overview}, we came up with an event-driven microservice-based architecture consisting of several microservices, RabbitMQ as the event-handling message broker, and MySQL as the internal database (IDB). Note that the components depicted in green are open-source tools, while the components depicted in grey were developed as part of this work. Moreover, we use a simple centralized logging facility through state of the art tools such as Elasticsearch and Fluentbit to lessen the burden of collecting log data from each service, and finally Kibana, to visualize the gathered logs.

\subsection{Components}

Figure~\ref{fig:arch-overview} depicts all the components as part of the proposed design. Their responsibilities are described below:

\paragraph{API Server} Responsible for receiving requests from users such as triggering a crawl etc., processing these tasks, emitting appropriate events to the message broker, and returning an appropriate response to the user. The \textit{curl} command and the Swagger~\footnote{Swagger, https://swagger.io/} user interface are two methods that users may use to interact with REST APIs in order, e.g., to add a repository for analysis. Alternatively, a daemon service may be used to call the API server regularly without user involvement. 

Whenever the API Server sends out a message, a routing key is attached to it, allowing other services to receive events based on such key. RabbitMQ is responsible for delivering messages to the correct consumers. In this sense, the API server works as an event producer for RabbitMQ, generating relevant events picked up later by the Pika library, we used to bind RabbitMQ consumers in Python. Besides that, a health endpoint is provided in order to determine the current state of the API server at any moment.

\paragraph{NVD Cloner and EPSS Cloner} Responsible for, obtaining the publicly accessible NVD and EPSS data, respectively, processing it, and storing it in the database. 

\paragraph{Daemon Service} Used to initiate common tasks without user intervention. For instance, the Daemon service triggers the recurrent NVD clone activity by making a simple API call to the server. Similarly, every hour it schedules analysis tasks performed by the Analyser which checks if the collected dependencies are affected by newly discovered vulnerabilities etc.

\paragraph{Github Crawler} Retrieves metadata from popular repositories in a variety of programming languages, using, e.g., the number of stargazers as the popularity metric.
We furthermore gather additional relevant metadata and ﬁelds, such as lines of code, lines of comments, number of ﬁles, number of commits, and number of contributors, using a distinct set of tools within a distinct worker-managed process, as described next.

\paragraph{Worker} Logical unit composed of the BOM Generator and the BOM Crawler services. The first can be initiated by the API server, and performs sophisticated processing steps to produce an SBOM for each release of a repository and transports the resulting SBOM to a shared location. We can then run the BOM Crawler using the API server to read the SBOM ﬁles and add the components to the database simultaneously. Once the components have been added to the internal database, the Analyser may be used to determine whether they are vulnerable or not. Note that all the processes are decoupled from one another, which paves the way for highly scalable system which allows to track several thousand of projects in parallel.

Depending on the total number of repositories and releases, several instances of workers can be spawned to generate SBOMs in parallel, since each worker’s BOM Generator is capable of receiving events in a round-robin method. Similarly, the API Server offers an additional API for simultaneously launching BOM Crawlers on all workers through a broadcast.

\paragraph{Component Analyser} Makes use of Sonatype OSS Index's public REST API to determine the vulnerable components of the given system and updates our internal database with this information, mapping the component to the identified vulnerability.

\paragraph{Internal Database} MySQL is a widely-used, open-source relational database management system. Using SQL queries, data storage and retrieval is quite simple. As a result, MySQL is used as database in our system. Data stored in the database includes, but not limited to, information on each component of \voda, the repositories gathered by the GitHub Crawler, their releases and commits and the vulnerabilities encountered with regards to the open source dependencies.

\section{Implementation} \label{sec:implementation}

In this section, we discuss the implementation details of all the services from our architecture as depicted in Figure~\ref{fig:arch-overview}. As for the implementation, we opted for Python as it allowed us to rapidely build our system in a memory safe manner. One of the libraries we used is \textit{Python Flask} for the creation of web applications, specially used to implement the API server, \textit{RabbitMQ} as a message broker to facilitate the transit of information between the services, and the \textit{Pika} librabry, which provides a simple interface for client programs to connect with RabbitMQ hosts. As for the management of the internal MySQL database, we utlized SQLAlchemy, a widely known ORM package for Python applications that offers high-level APIs for working with relational databases. Besides that, to promote a distributed, easy to manage and scalable system, all the services are deployed as Docker containers and orchestrated by a Kubernetes cluster which allows us to seamlessly scale the system by defining the number of replicas for each service. 

\subsection{Cloning vulnerability datasets} \label{subsec:implementation:nvd}

In this work, we chose the National Vulnerability Database (NVD) and the Exploit Prediction Scoring System (EPSS) as the datasets of interest. From the NVD, we acquire data regarding disclosed vulnerabilities and their severity, and from EPSS, we collect data that indicates the probability that a vulnerability may be exploited. In our approach, the NVD Cloner and the EPSS Cloner are the two components responsible for handling this information.

The NVD Cloner acts as a RabbitMQ consumer and was implemented using the Pika library, which receives all messages issued by the API Server with the same routing key used upon the consumer startup. Upon receiving a message, the NVD Cloner downloads the NVD data feed from $2002$ to $2023$ and classifies it into \textit{annual} and \textit{modified} data feeds. The \textit{annual} data feed remains constant while the \textit{modified} one is updated on a regular basis. After downloading, the NVD Cloner extracts the files containing all vulnerability information and stores it in the internal database. The NVD data feed, in its turn, is stored locally, and when a subsequent request is made to the NVD Cloner, it will identify the local copy and avoid repeating the process for the \textit{annual} data feed. Instead, it will only attempt to update the \textit{modified} one in order to maintain the NVD data updated in the internal database.

The workflow for the EPSS Cloner is very similar to the one described above. This component is also implemented as a RabbitMQ consumer and follows the same procedure as the NVD Cloner in order to receive new messages. The EPSS Cloner then acquires the EPSS data feed, and stores the extracted information in the internal database. This data includes an identifier for the vulnerability, a percentile value and a score, which shows the likelihood that a vulnerability will be exploited.

\subsection{Mining software repositories} \label{subsec:implementation:mining}

The GitHub Crawler is developed as a RabbitMQ consumer that gathers information about open-source GitHub repositories, following the same procedure as other components that exchange data via the RabbitMQ broker service. In order to gather this information, different approaches are necessary since part of the metadata about the given repositories, such as the number of files, lines of code, lines of comments, contributors, and commits, for each release, cannot be obtained using just the GitHub REST APIs. Thus, we split this process into two stages:

\paragraph{Stage 1} Initially, the GitHub Crawler captures basic information about a given repository, such as its name, clone URL and programming language, by using the GitHub API. The API also provides IDs for the repository and its releases, which we persist in the internal database such that in future executions of the GitHub Crawler, we are able to determine if a repository is already in our internal database or not, and can proceed to only update the repository through a simple pull in order to retrieve new releases.

\paragraph{Stage 2} Once we have gathered the basic information of each repository, the system moves to the next stage. In this stage, the API Server retrieves all repositories from our database and sends the repository ID to the RabbitMQ service as the message payload. Then, worker instances of the BOM Generator consume such messages, and run asynchronously on distinct repositories, perform checkouts of every release tag, and extract the missing metadata regarding the number of lines of code, files, commits, and contributors. Note that we keep only repositories in our database that have at least a single release (tag).

\subsection{Language agnostic SBOM generation} \label{subsec:implementation:state-machine}

Our system spawns a new instance of the BOM generator for each repository that is retrieved from the database which allow us to analyze several repositories and projects in parallel.
The BOM generator then clones the repository, identifies its programming language, and then attempts to generate an SBOM file for each release tag of that repository. 

Since several tools are provided by CycloneDX to facilitate SBOM generation as discussed in Section~\ref{subsec:background:sbom}, we need to invoke the appropriate CycloneDX tool to generate the SBOM.
In order to do that, we implemented a state machine based approach that executes different tools depending on the identified programming language and, as a result, creates the expected SBOM. 

To build our state machine we defined three main states named \textit{Init}, \textit{BOM Generation}, and \textit{Cleanup}, and dynamically generate the relevant states for the specific programming languages of each repository.
The state machine initiates in the \textit{Init} state, transitions to all other states based on the transition rules, and ultimately reaches the \textit{Cleanup} state. Note that this adaptable architecture enables us to easily add support for more programming languages in the future. 

As an example, we briefly outline the state machine that generates an SBOM for a Go repository. For each release, we build a state machine containing states that are written particularly for a Go workflow: 

\paragraph{Init state} In this state, the system checks out the release tag and then verifies if a \texttt{go.mod} file exists in the root directory of the repository. If it exists, it transitions immediately to the \textit{BOM Generation} state. If the \texttt{go.mod} file is absent, it will try to dynamically create one using the \texttt{go} CLI tool. If this step fails, it will transition to the \textit{Cleanup} state since the \texttt{go.mod} file is a requirement for the CycloneDX Go Plugin to successfully create a SBOM.

\paragraph{BOM Generation state} Since the application at hand is a Go application, the state machine will use the CycloneDX Go Plugin. Hence, the \texttt{go.mod} file will be passed to the tool in order to generate the SBOM file. The generated SBOM file contains all the components of the given application and its dependencies. After the SBOM generation, the state machine will transition to the \textit{Cleanup} state.

\paragraph{Cleanup state} If there is no SBOM file present, the state machine will update the release state of the particular project/repository in the database and set it to \texttt{FAIL}. If an SBOM file was successfully created, the release state will be set to \texttt{DONE} preventing a reprocessing of the same release.
The ready to use SBOM file will then be transferred to a shared directory which is also accessible by the BOM Crawler. If the \texttt{go.mod} file was produced as part of the \textit{Init} state, it will be deleted.
The system will finally execute a \texttt{git reset} to delete all temporary modifications made to the release and to remove the created SBOM file from the repository.

\subsection{SBOM analysis} \label{subsec:implementation:sbom-analysis}

For the following analysis of the SBOM files, we opt to use the Sonatype OSS Index as it offers an endpoint that accepts a list of Package URLs (PURL) and returns the corresponding vulnerabilities for each dependency. However, in order to avoid overloading Sonatype's public API as we are processing several repositories in parallel, we implemented a two-step SBOM analysis as follows:

In this two-step approach, each BOM Crawler reads an SBOM file and adds newly discovered package information to the database and updates its status to \texttt{NEW}. Hence, packages/dependencies discovered by parallel instances will be omitted.
Moreover, we also implemented a recursive algorithm that calculates a dependency depth from the dependency graph stored in the SBOM file in order to also register transitive dependencies that way. 
        
In the second step, the Component Analyser retrieves all components with the state \texttt{NEW} from the database, and uses Sonatype OSS Index to identify the ones that are vulnerable.
Each event received by the Component Analyser attempts to examine $500$ components from the database, in chunks of $25$ components for every Sonatype REST request.
After successful retrieving the information, the internal database will be updated with the mapping between the found vulnerability and the respective component.




\section{Evaluation} \label{sec:evaluation}

\subsection{Environment and methodology}

For our evaluation, we deployed our system on a $8$ nodes Kubernetes cluster, each equipped with $8GB$ of main memory and a $8$ core CPU. The cluster runs all components as depicted in Figure~\ref{fig:arch-overview} where we either used helm~\cite{K8sHelm} (the package/deployment manager for Kubernetes) to deploy components or a dedicated operator such as for RabbitMQ~\cite{RabbitmqOperator}. For data persistence, we opted to use a MySQL instance as our internal database.

\subsection{Quantitative analysis of open source repositories}

In order to gain more insights about how wide spread vulnerabilites are, and for how long they persist in repositories in general, we collected data from $1042$ public open-source repositories hosted on GitHub which translates to a total of $49055$ releases.
Repositories without any releases were excluded from our dataset and the popularity was measured considering its number of stargazers. Note that the absolute number of repositories varies between programming languages. 
A low number of repositories were excluded due to either version incompatibility, missing dependencies, or a very project-speciﬁc setup which did not allow us produce proper SBOMs for some of the releases.
However, the dataset contains $841$ repositories where it was possible to generate SBOMs with a total of $34990$ releases where $90\%$ of these releases use Rust and Go as programming languages. 

\begin{figure}[t]
    \centering
    \begin{subfigure}[t]{0.24\textwidth}
        \includegraphics[width=\textwidth]{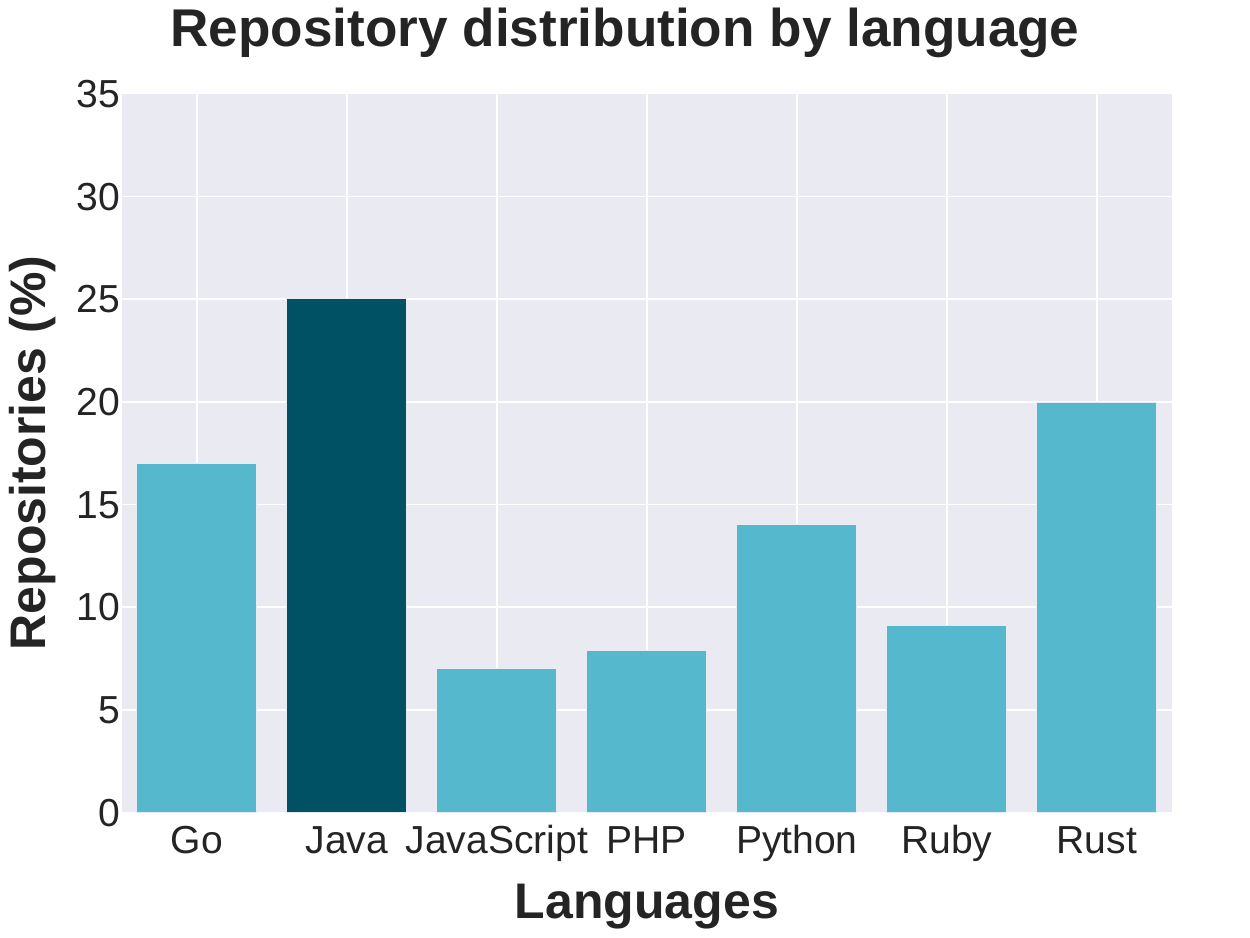}
        \caption{Repository distribution.}
        \label{fig:repository-dist}
    \end{subfigure}
    \begin{subfigure}[t]{0.24\textwidth}
        \includegraphics[width=\textwidth]{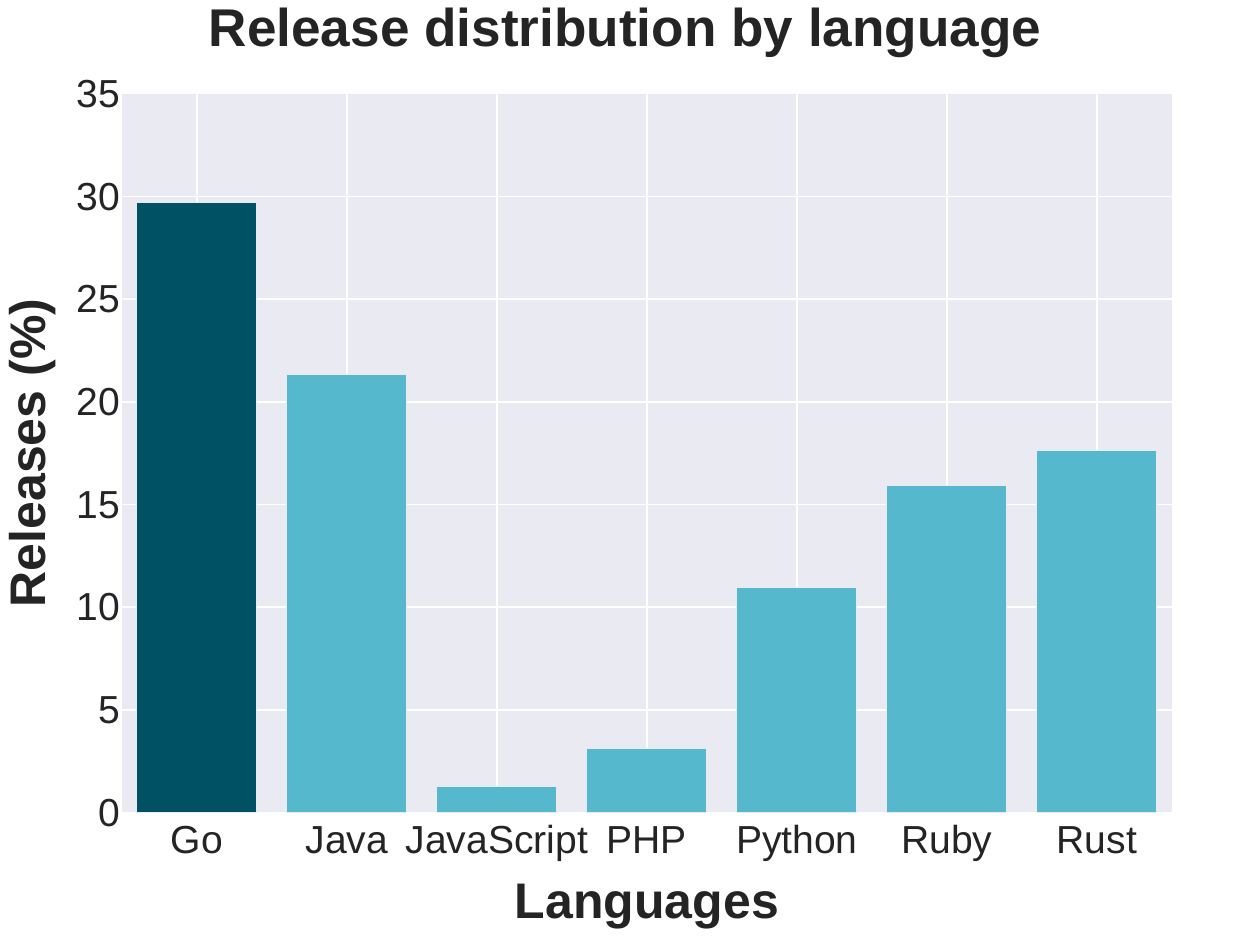}
        \caption{Release distribution.}
        \label{fig:release-dist}
    \end{subfigure}
    \caption{Repository and release distribution by programming language.}
    \label{fig:planguages-dist}
    \vspace{-1.5em}
\end{figure}

Figure~\ref{fig:planguages-dist} depicts the overall distribution of programming languages (a) as well as the number of releases (b) for each of the languages: Java, Go, Rust, Ruby, Python, PHP and JavaScript.
As shown in the Figure, the majority of the repositories uses Java ($25\%$), followed by Rust ($20\%$) and Go ($17\%$), which also applies to the release distribution except in a slightly different order for Go ($29.7\%$) being the programming language with the most releases.
Also, a deeper investigation reveals that some repositories date back to as early as 2013, indicating that the considered releases have been made between 2013 and 2023.
Hence, using the historic information, we can review a potential trend for the aforementioned programming languages throughout the past 10 years.

\begin{figure}[t]
    \centering
    \includegraphics[width=0.4\textwidth]{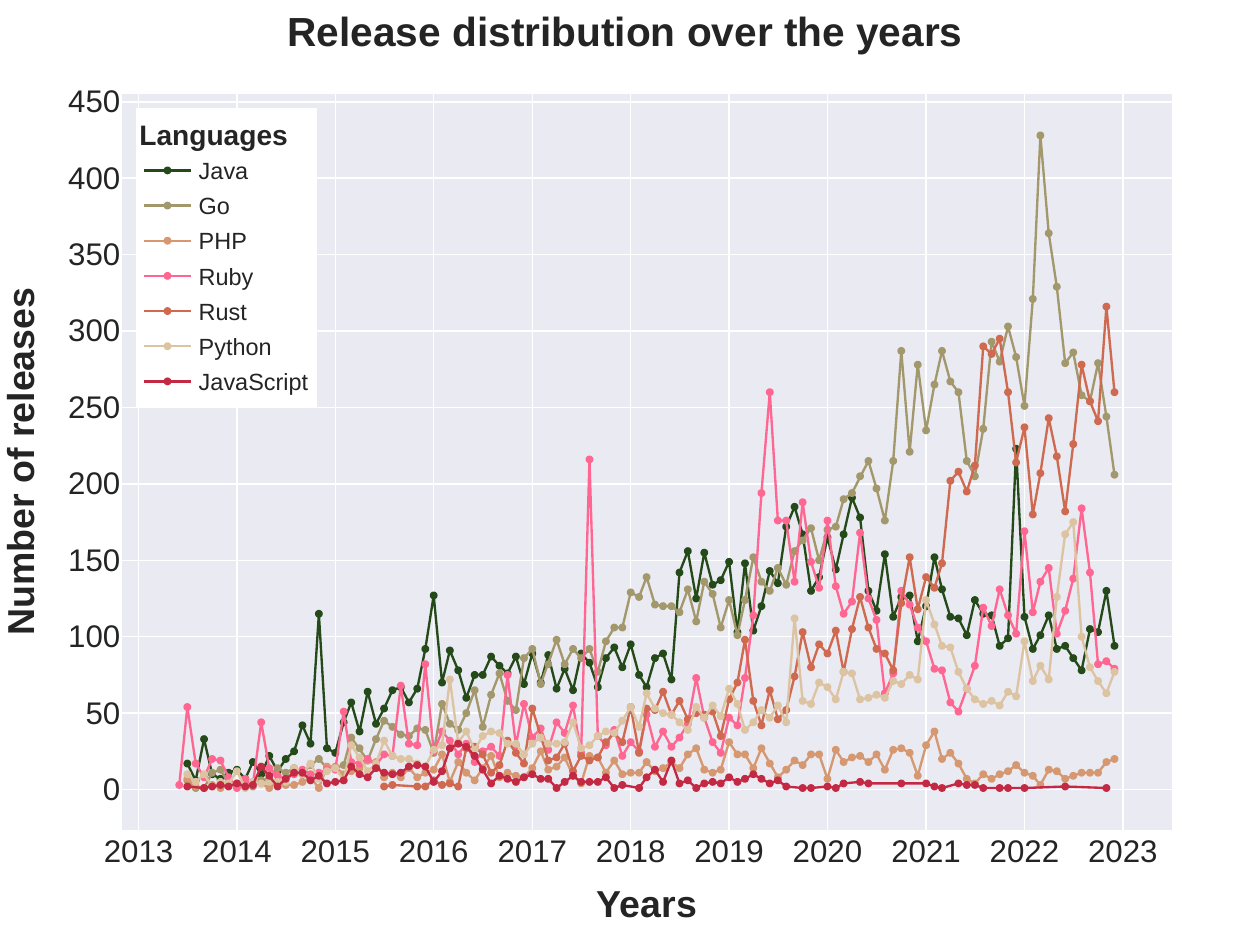}
    \caption{Number of releases over time.}
    \label{fig:release-timeline}
\end{figure}

Figure~\ref{fig:release-timeline} depicts the release pattern over time for each programming language. Notably, when compared to other programming languages, Go and Rust exhibit a strong upward trend in their release numbers.
Although there is a high number of releases overall for Java and Ruby, we can see that both languages fluctuate considerably over time, with no clear pattern that indicates growth, but rather a slight descend. Python, on the other hand, although having a smaller total number of releases, shows an upward trend throughout the years. From these findings we can conclude that Go, Rust, and Python are increasingly popular among developers in the open-source community.

\begin{figure}[t]
    \centering
    \begin{subfigure}[t]{0.24\textwidth}
        \includegraphics[width=\textwidth]{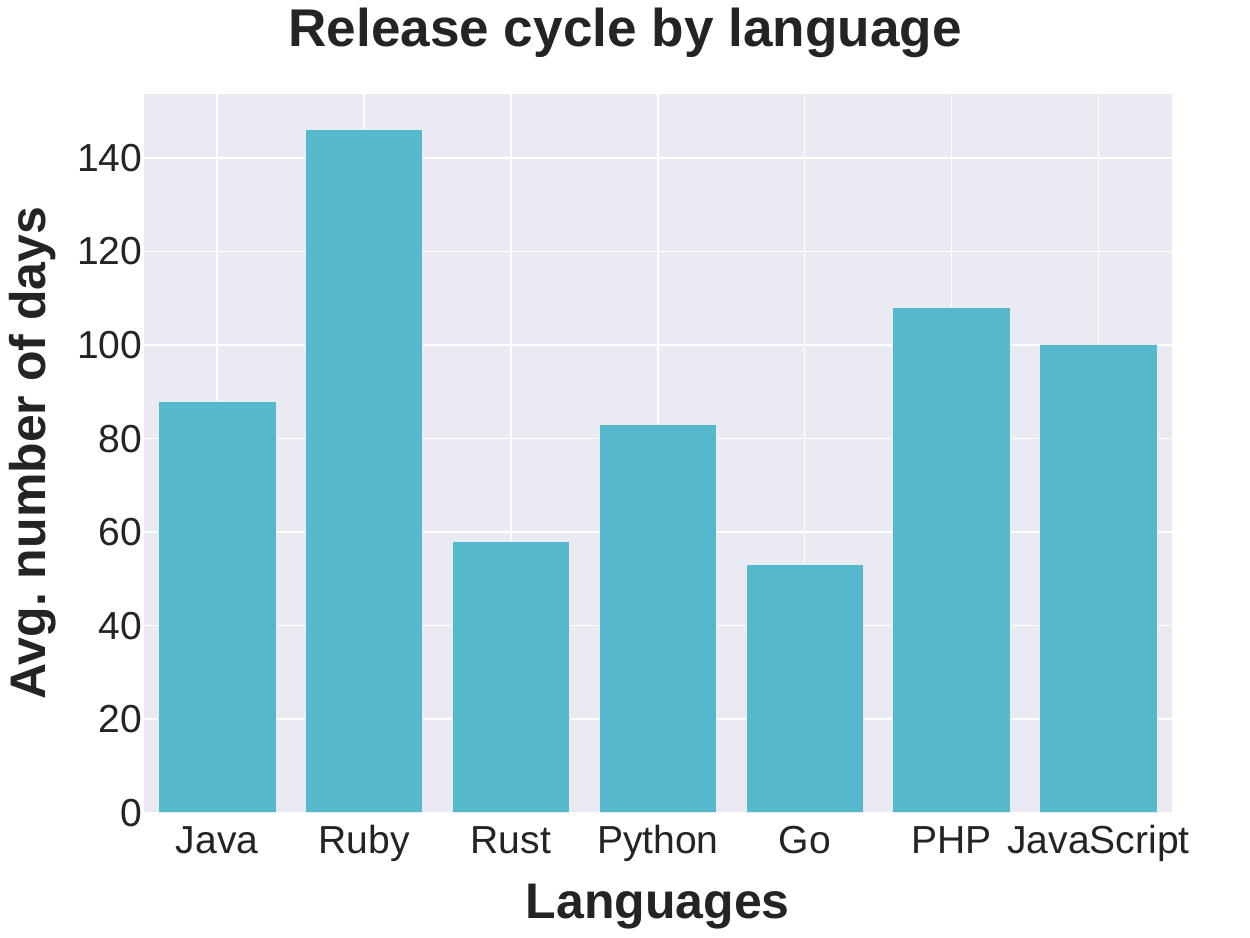}
        \caption{Average days for a release.}
        \label{fig:release-days-dist}
    \end{subfigure}
    \begin{subfigure}[t]{0.24\textwidth}
        \includegraphics[width=\textwidth]{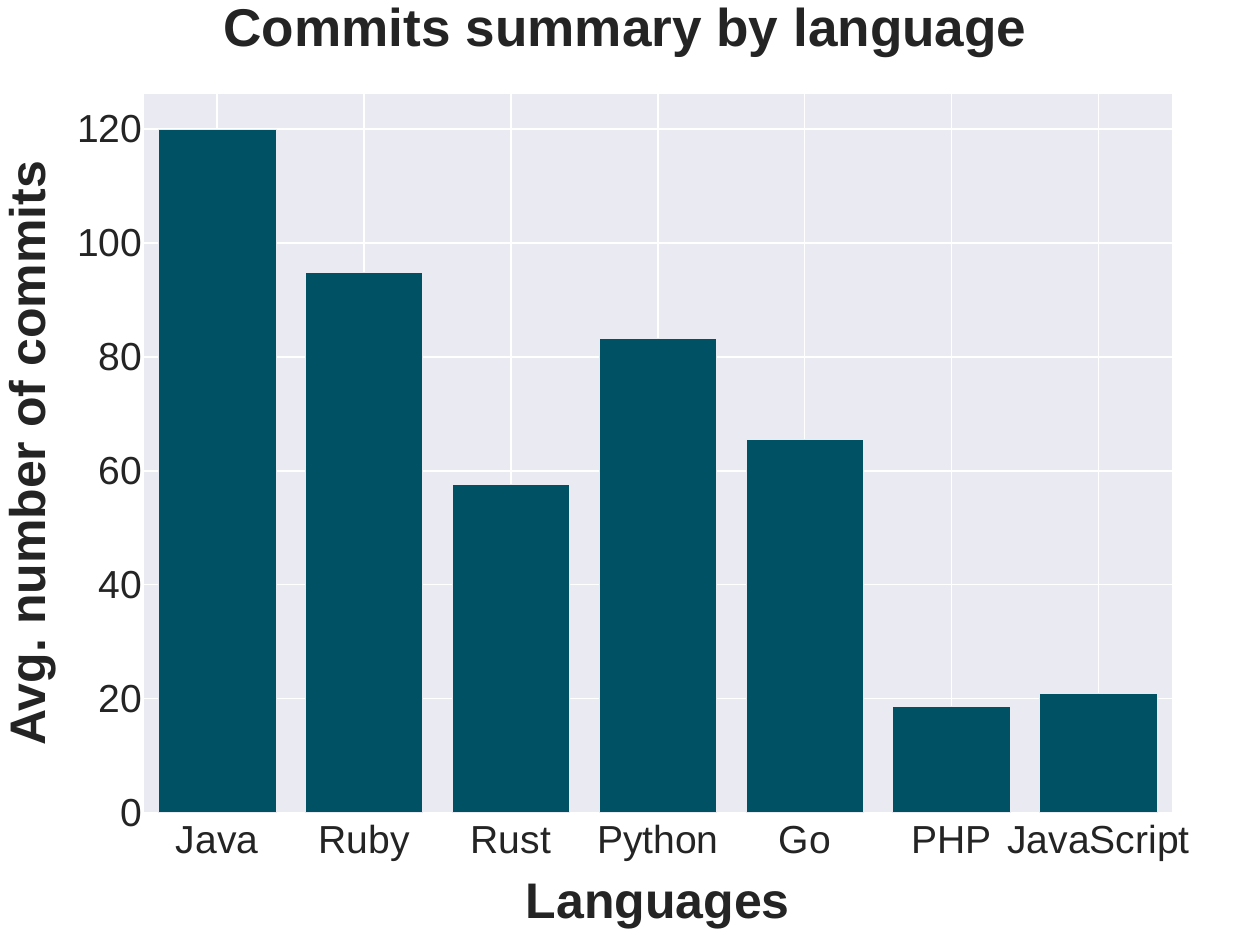}
        \caption{Average number of commits.}
        \label{fig:commits-dist}
    \end{subfigure}
    \caption{Release cycle and commits distribution by programming language.}
    \label{fig:release-cycle-dist}
    \vspace{-1.5em}
\end{figure}

Figure~\ref{fig:release-cycle-dist} shows the average number of days and commits for each release. As depicted in the Figure, we can see that the average number of new commits per release across all programming languages is around $60$ as shown in Figure~\ref{fig:commits-dist} (except for some outliers such as Ruby, Php and Javascript), and the average days for a release is around $80$ as shown in Figure~\ref{fig:release-days-dist}. From this, we conclude that Go and Rust are the language with more frequent releases in average, whilst still maintaining the average of commits per release.  

\subsection{Directive and transitive dependencies: depth comparison}

\begin{figure}[t]
    \centering
    \begin{subfigure}[t]{0.24\textwidth}
        \includegraphics[width=\textwidth]{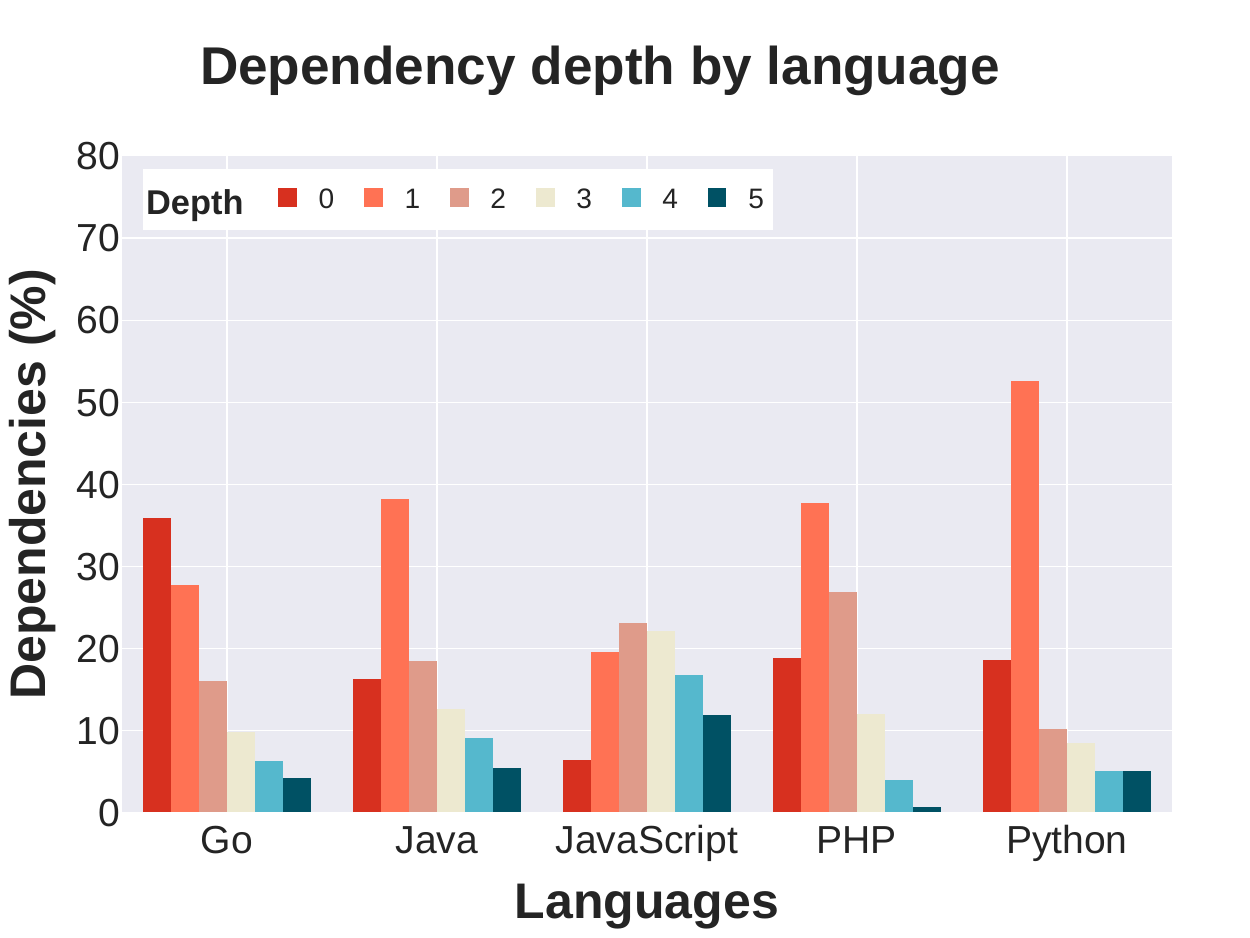}
        \caption{Dependency depth.}
        \label{fig:repository-dist}
    \end{subfigure}
    \begin{subfigure}[t]{0.24\textwidth}
        \includegraphics[width=\textwidth]{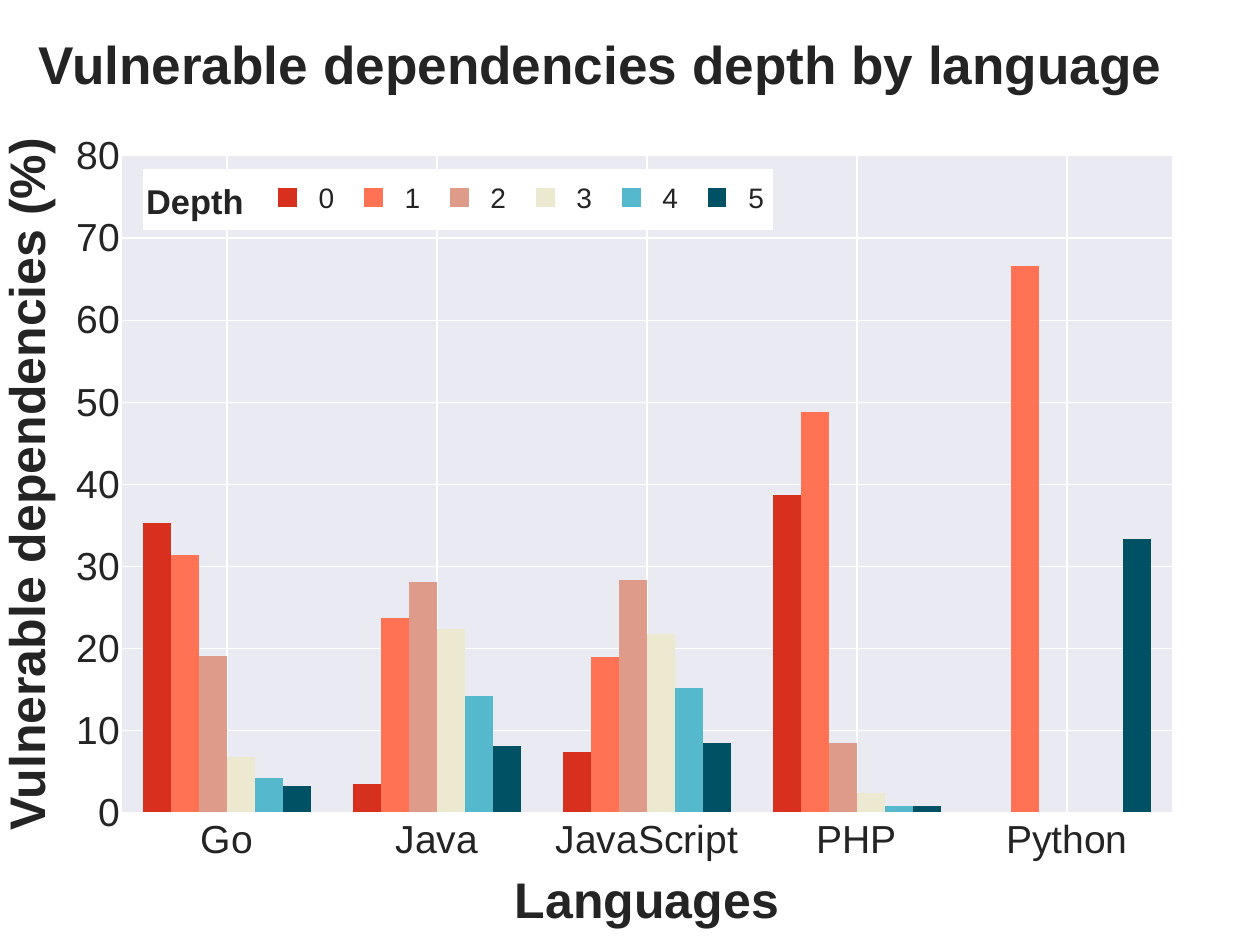}
        \caption{Vulnerable dependency depth.}
        \label{fig:release-dist}
    \end{subfigure}
    \caption{Dependency depth by programming language.}
    \label{fig:depth-relationship}
    \vspace{-1.5em}
\end{figure}

We furthermore investigated the dependencies in repositories with regards to their depth in the dependency tree, ranging from $0$ to $5$, for Java, Go, PHP, JavaScript, and Python.
Components with a depth in the tree of $0$, i.e., at the root are direct dependencies, whereas the remaining components are considered as transitive dependencies. The depth level increases from $1$ to $5$ as the chain of transitive dependencies increases. Figure~\ref{fig:depth-relationship} depicts the distribution dependencies in terms of depth level and programming language.
From this, we can observe that Go has a higher percentage of direct dependencies in comparison to other programming languages, while transitive dependencies of depth level $1$ are the most common among all languages. 

As for components that contain vulnerabilities, we can see that Go and PHP have a similar higher number of direct non-transitive vulnerable dependencies.
Java, on the other hand, has a smaller number of vulnerable direct dependencies and is well distributed in terms of transitive dependencies, ranging from $1$ to $5$ depth levels. We can also conclude that in general, vulnerabilities are mostly concentrated in transitive dependencies, specially in Java projects. 

\subsection{Case study}

In this section, we further analyze how the release patterns evolved over the past ten years for Java and Go releases, from 2013 to 2023 based on several incidents.
For each of the programming languages mentioned, the release pattern is examined, along with an example vulnerability case study, to determine if any changes have occurred in the release pattern.
Subsequently, the findings on the correlation between commits, contributors, and the number of vulnerabilities is presented.

\begin{figure}[t]
    \centering
    \begin{subfigure}[t]{0.24\textwidth}
        \includegraphics[width=\textwidth]{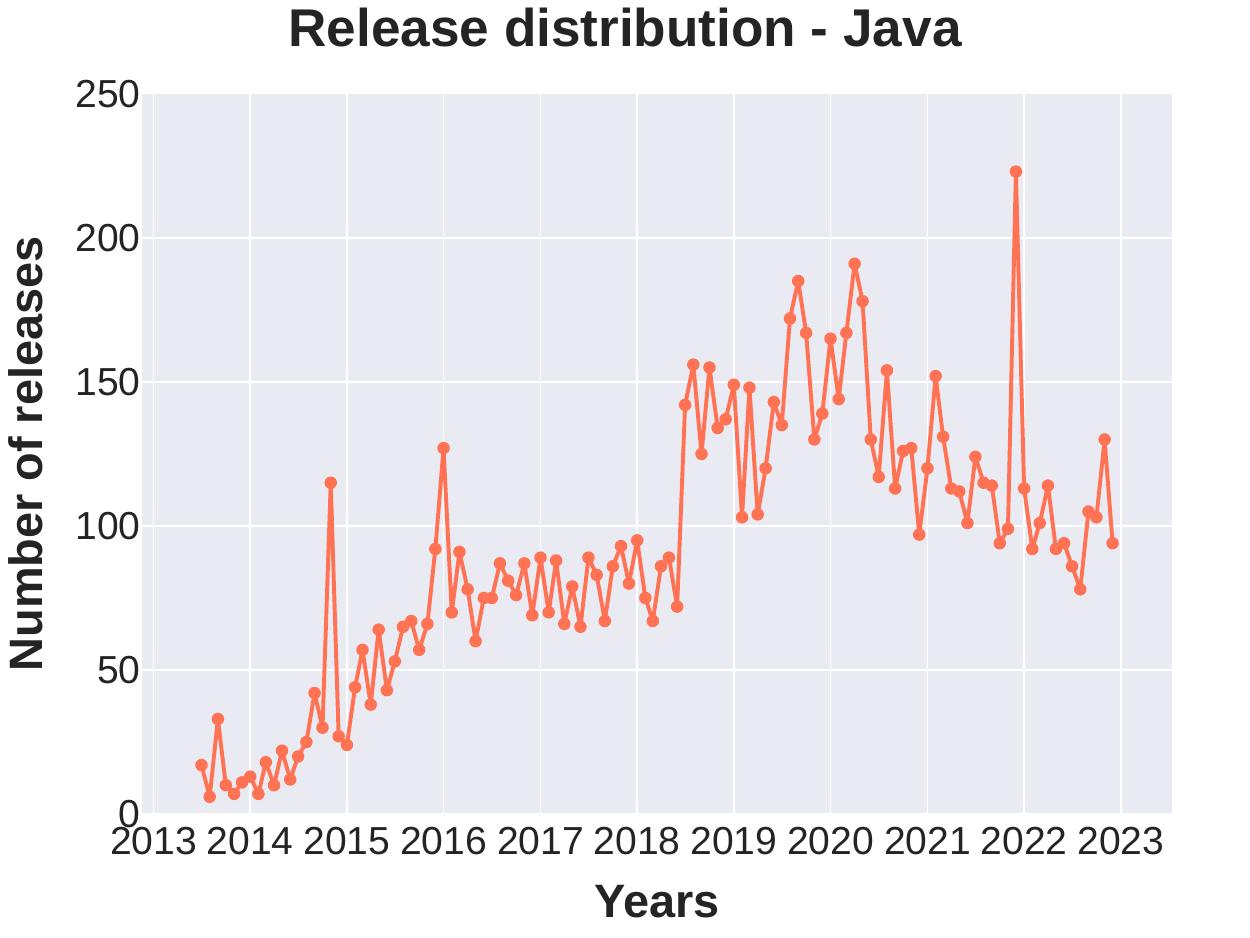}
        \caption{Java releases.}
        \label{fig:release-java}
    \end{subfigure}
    \begin{subfigure}[t]{0.24\textwidth}
        \includegraphics[width=\textwidth]{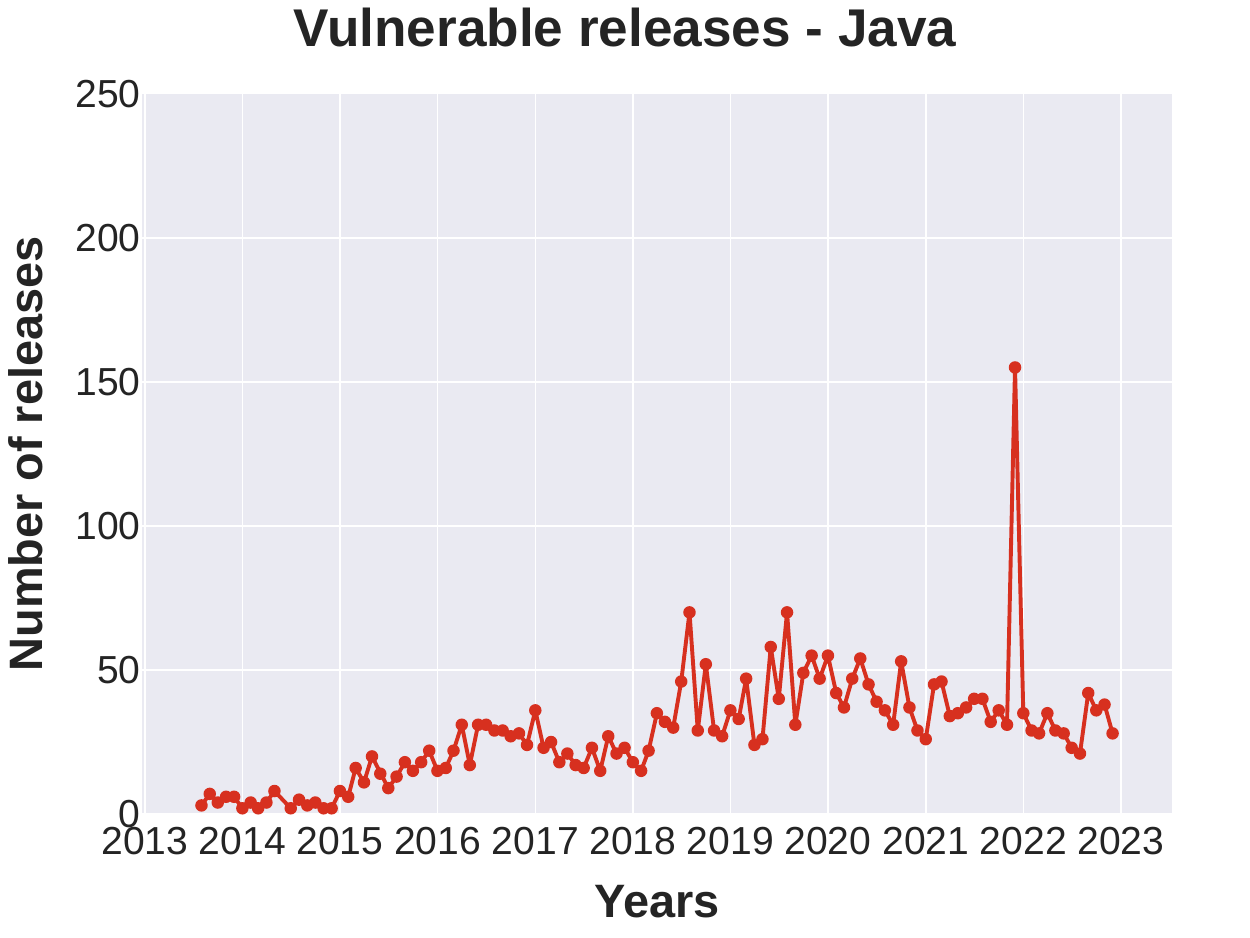}
        \caption{Java vulnerable releases.}
        \label{fig:release-vuln-java}
    \end{subfigure}
    \caption{Release pattern - Java.}
    \label{fig:release-pattern-java}
    \vspace{-1.5em}
\end{figure}

Figure~\ref{fig:release-java} depicts a consistent growth in the annual Java release count until 2020, followed by a slight decline. Nevertheless, it is noteworthy that there has been a resurgence in the number of releases towards the conclusion of 2021. In Figure~\ref{fig:release-vuln-java}, we observe the quantity of releases that encompassed at least one vulnerability over the recent years, with a significant surge in vulnerable releases occurring around the end of 2021.

\begin{figure}[t]
    \centering
    \includegraphics[width=0.35\textwidth]{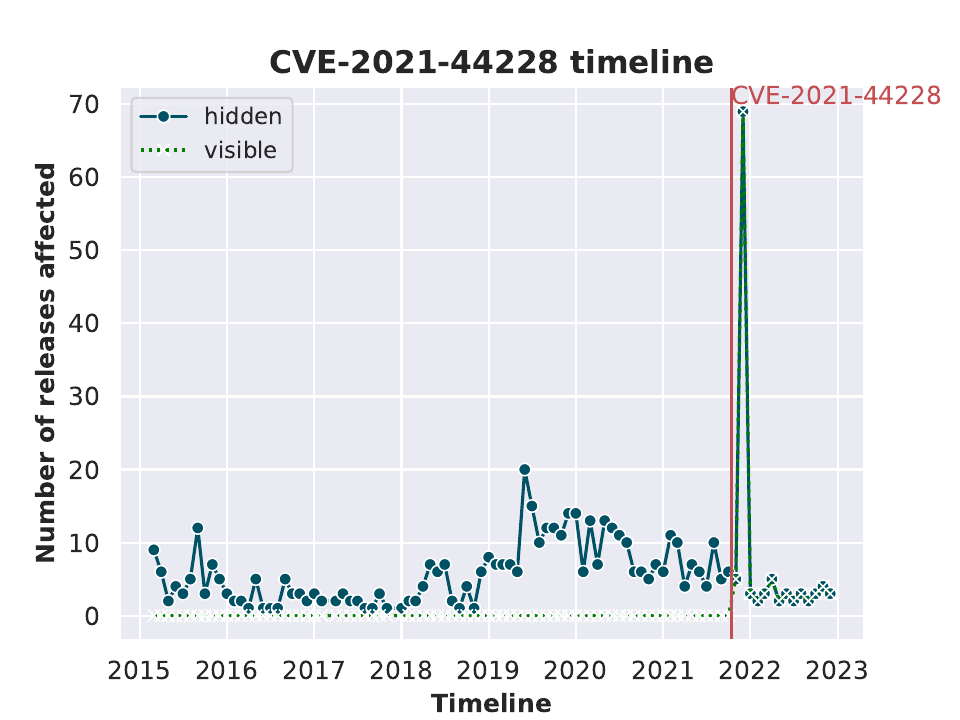}
    \caption{The Log4Shell zero-day vulnerability.}
    \label{fig:cve-log4shell}
    \vspace{-1.5em}
\end{figure}

This behavior can be explained by the disclosure of the Log4Shell vulnerability. It possesses a CVSS $v3$ score of $10$, which represents the highest possible value for a vulnerability, classified as \textit{CRITICAL}. Additionally, it has an EPSS score of $0.97095$, indicating a high probability of exploitation. Upon examining the affected releases of the Log4Shell vulnerability, it is evident that despite its discovery on December 2021, numerous previous versions were already vulnerable. Furthermore, Figure~\ref{fig:cve-log4shell} illustrates a subsequent increase in Java releases followed by a decline after the disclosure of the vulnerability.

\begin{figure}[t]
    \centering
    \begin{subfigure}[t]{0.24\textwidth}
        \includegraphics[width=\textwidth]{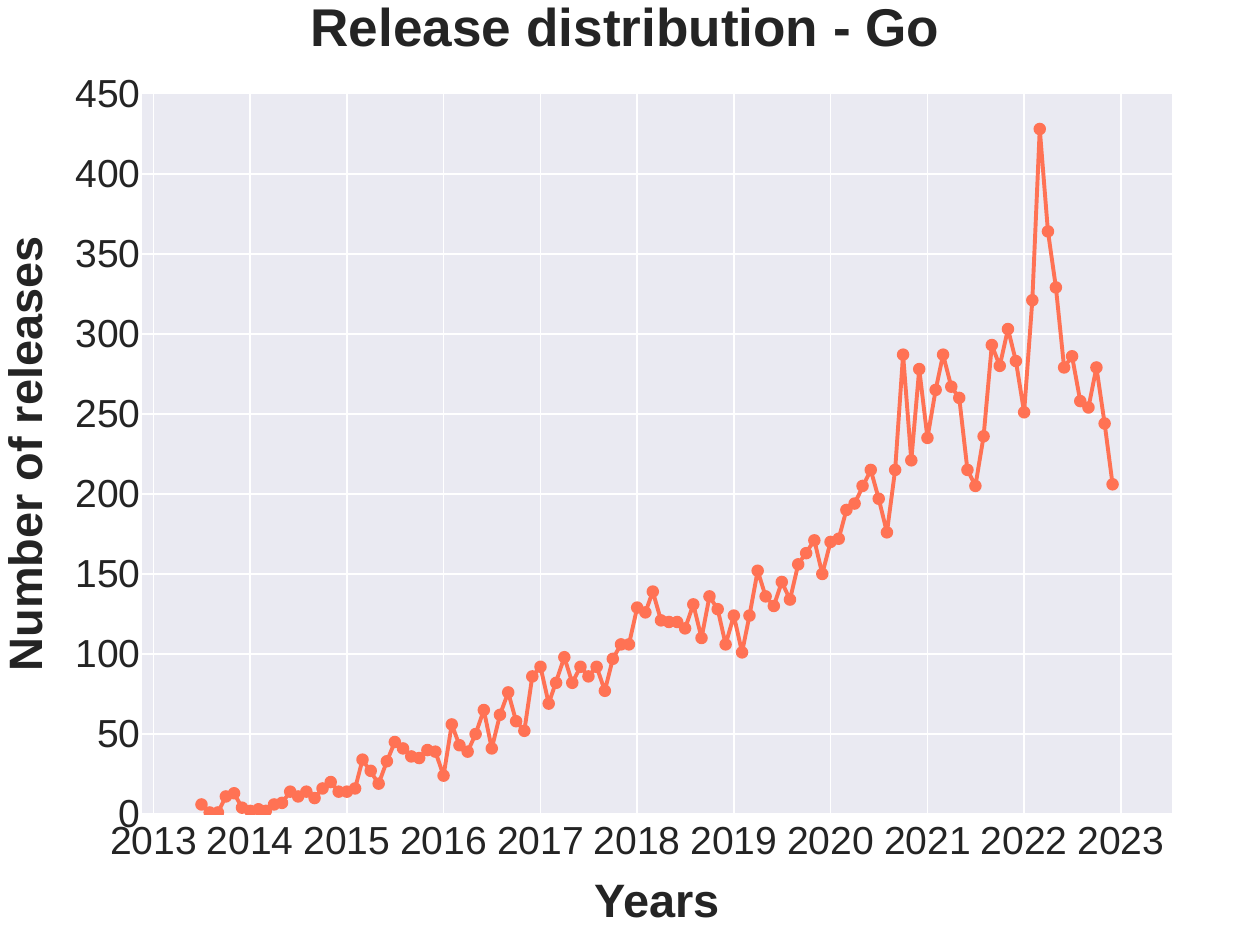}
        \caption{Go releases.}
        \label{fig:release-go}
    \end{subfigure}
    \begin{subfigure}[t]{0.24\textwidth}
        \includegraphics[width=\textwidth]{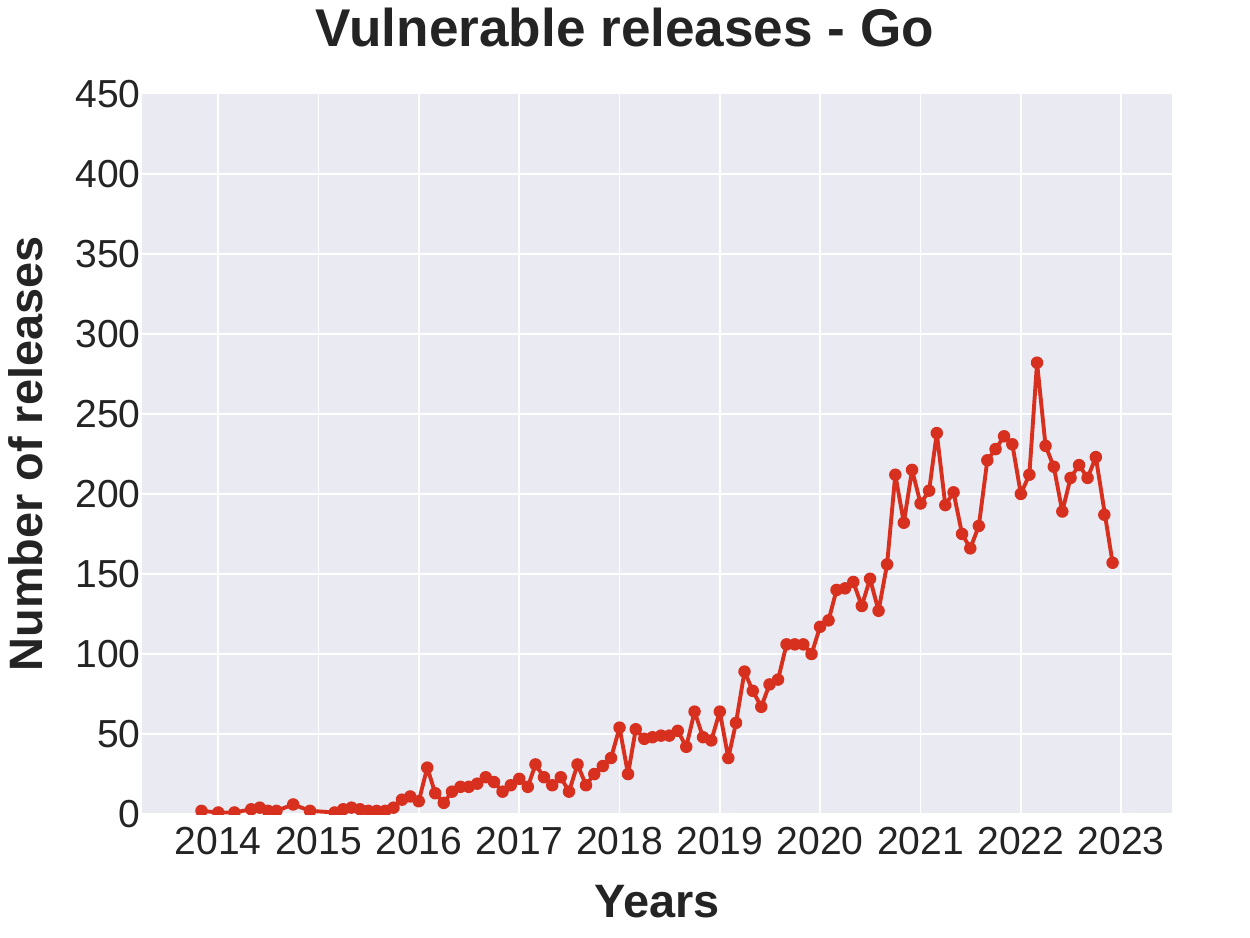}
        \caption{Go vulnerable releases.}
        \label{fig:release-vuln-go}
    \end{subfigure}
    \caption{Release pattern - Go.}
    \label{fig:release-pattern-go}
    \vspace{-1.5em}
\end{figure}

As for Golang, the annual releases have steadily increased until about 2022 and declined in the recent years, as depicted in Figure~\ref{fig:release-go}. Furthermore, Figure~\ref{fig:release-vuln-go} illustrates that the number of releases containing vulnerabilities is also on the rise. Additionally, our investigation indicates that Go repositories have a higher count of stargazers compared to repositories of other programming languages. This information combined indicates that there is a growing interest in Golang among developers and companies, resulting in more people using it. Go's popularity as a programming language can be attributed to its ease of use, execution speed, and excellent memory management. It is also worth noting that Go is utilized for developing internal tools for several big tech companies, including Google.

\begin{figure}[t]
    \centering
    \includegraphics[width=0.35\textwidth]{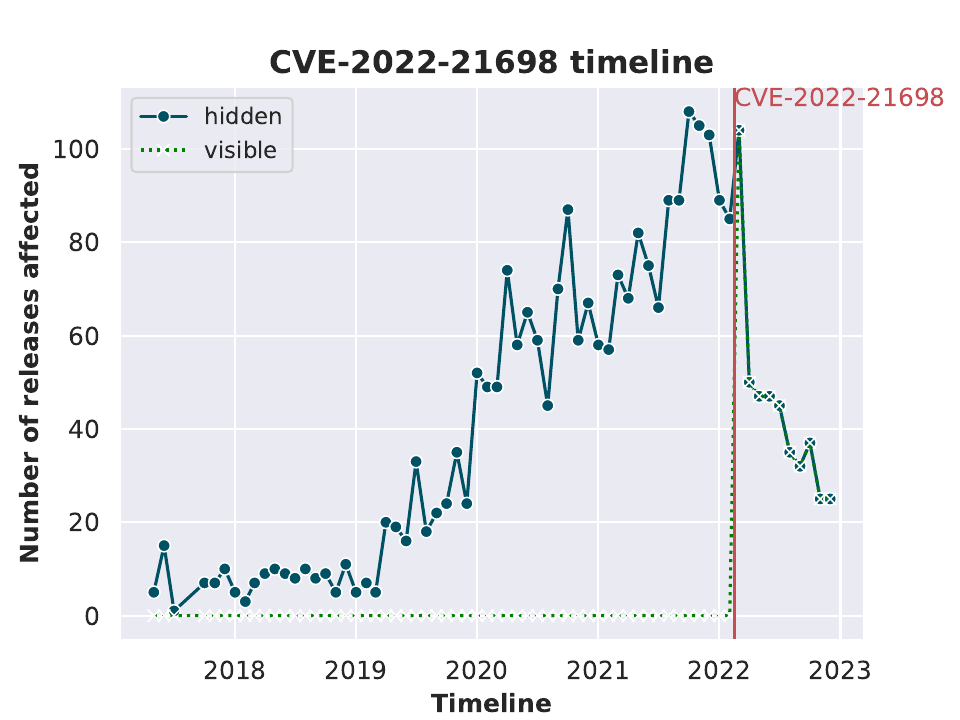}
    \caption{The client\_golang incident.}
    \label{fig:cve-client-golang}
    \vspace{-1.5em}
\end{figure}

The Go library known as \texttt{client\_golang} has been specifically developed to be used in conjunction with Prometheus, a widely known monitoring tool. A number of versions of this library were found to be impacted by a vulnerability identified as CVE-2022-21698~\cite{CVE-2022-21698}, which was announced on February 2022.
Out of the $48$ versions present in our database, $22$ were found to be susceptible to this particular vulnerability. The CVSS score assigned to it is $7.5$, while the EPSS score stands at $0.02686$. Despite the severity of the vulnerability being classified as \textit{HIGH}, the likelihood of exploitation is relatively low due to the lower EPSS score. Figure~\ref{fig:cve-client-golang} illustrates a decline in the number of releases affected by this vulnerability subsequent to its disclosure.

\subsection{Correlation analysis}

In this section, we try to identify correlations such as if more commits result in a higher amount of vulnerabilities which is driven by the intuition that an active repository/project which is quickly evolving may introduce more vulnerabilities compared to a dead project which has only maintenance releases where each commit only fixes security issues rather than bringing in new features and potentially new libraries with vulnerabilities.

Hence, in addition to the analysis on how incidents impact releases and release cycles, we investigated if properties such as number of commits, contributors and others have an impact on vulnerability injection as well as removal from projects. 
Table~\ref{table:vulnerabilit_attribute_correlation} presents the results obtained through the utilization of the Pearson correlation coefficient, including the results for a few more programming languages. 

In general, a moderately positive correlation is observed between the number of commits and the number of vulnerabilities in Go and Rust, while a moderate to weak positive correlation is found between their number of contributors and the number of vulnerabilities. Conversely, a moderately negative correlation is identified in PHP and Ruby. No statistically significant associations are detected in the case of Java, JavaScript, and Python. Consequently, we can conclude that in the context of Go and Rust, a higher number of commits corresponds to an increased number of vulnerabilities. However, this relationship does not hold for Java, PHP, and Ruby. It is important to note that the presence of inconsistent findings across different programming languages prevents us from assuming that a greater number of commits invariably leads to a higher number of vulnerabilities.



\begin{table}[t]
    \caption{Correlation between attributes and vulnerabilities}
    \centering
    \begin{tabular}{c|c|c}
	& Commits vs. Vuln. & Contributors vs Vuln. \\
	\hline
	Go & 0.350 & 0.394 \\
	Java & -0.076 & -0.082 \\
	JavaScript & 0.040 & -0.048 \\
	PHP & -0.301 & -0.219 \\
	Python & 0.033 & 0.199 \\
	Ruby & -0.274 & -0.328 \\
	Rust & 0.493 & 0.176 \\
    \end{tabular}
    \label{table:vulnerabilit_attribute_correlation}
\end{table}

\subsection{Persistence and occurrence of vulnerabilities}

\begin{figure}[t]
    \centering
    \begin{subfigure}[t]{0.24\textwidth}
        \includegraphics[width=\textwidth]{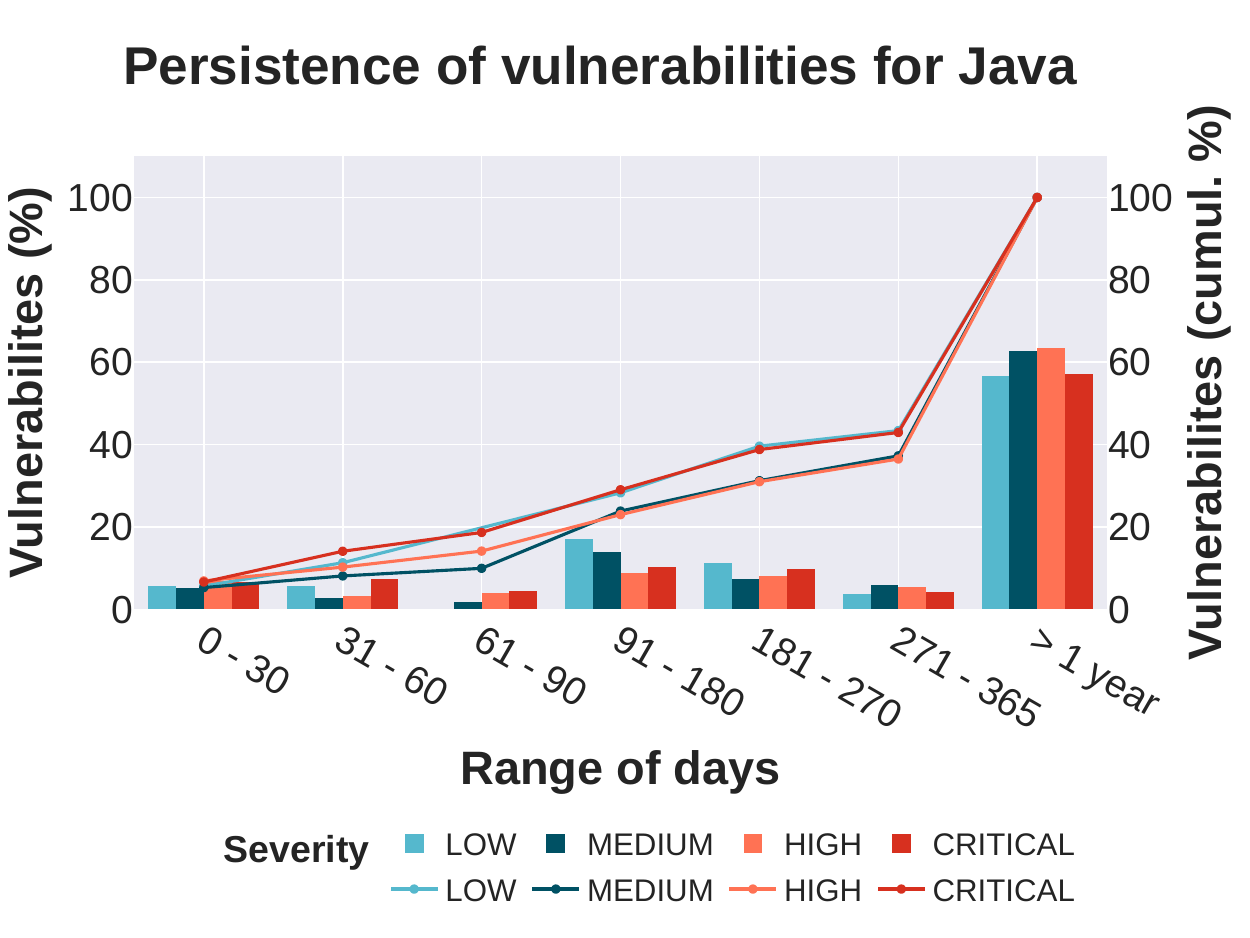}
        \caption{Java.}
        \label{fig:persistence-java}
    \end{subfigure}
    \begin{subfigure}[t]{0.24\textwidth}
        \includegraphics[width=\textwidth]{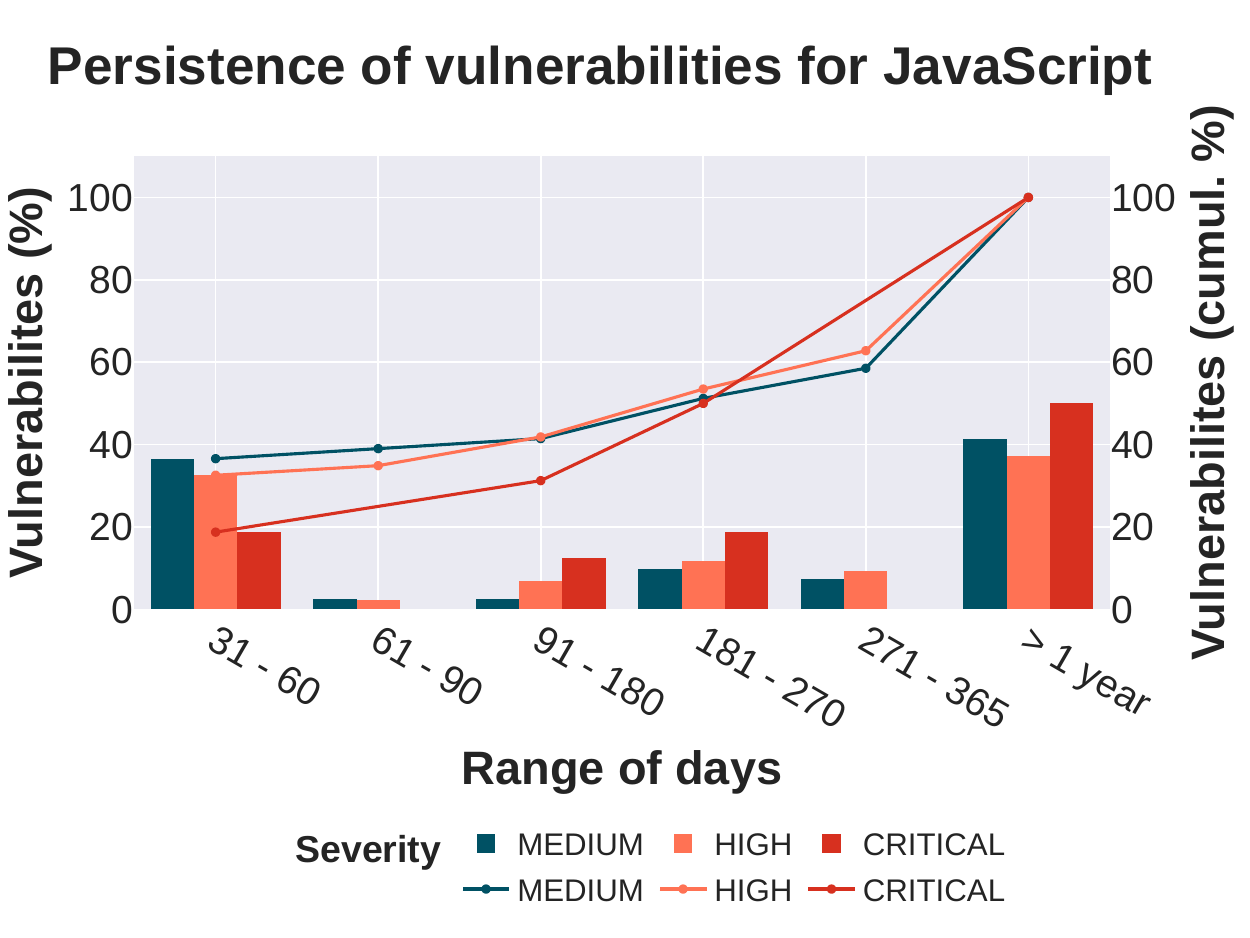}
        \caption{JavaScript.}
        \label{fig:persistence-javascript}
    \end{subfigure}
    \begin{subfigure}[t]{0.24\textwidth}
        \includegraphics[width=\textwidth]{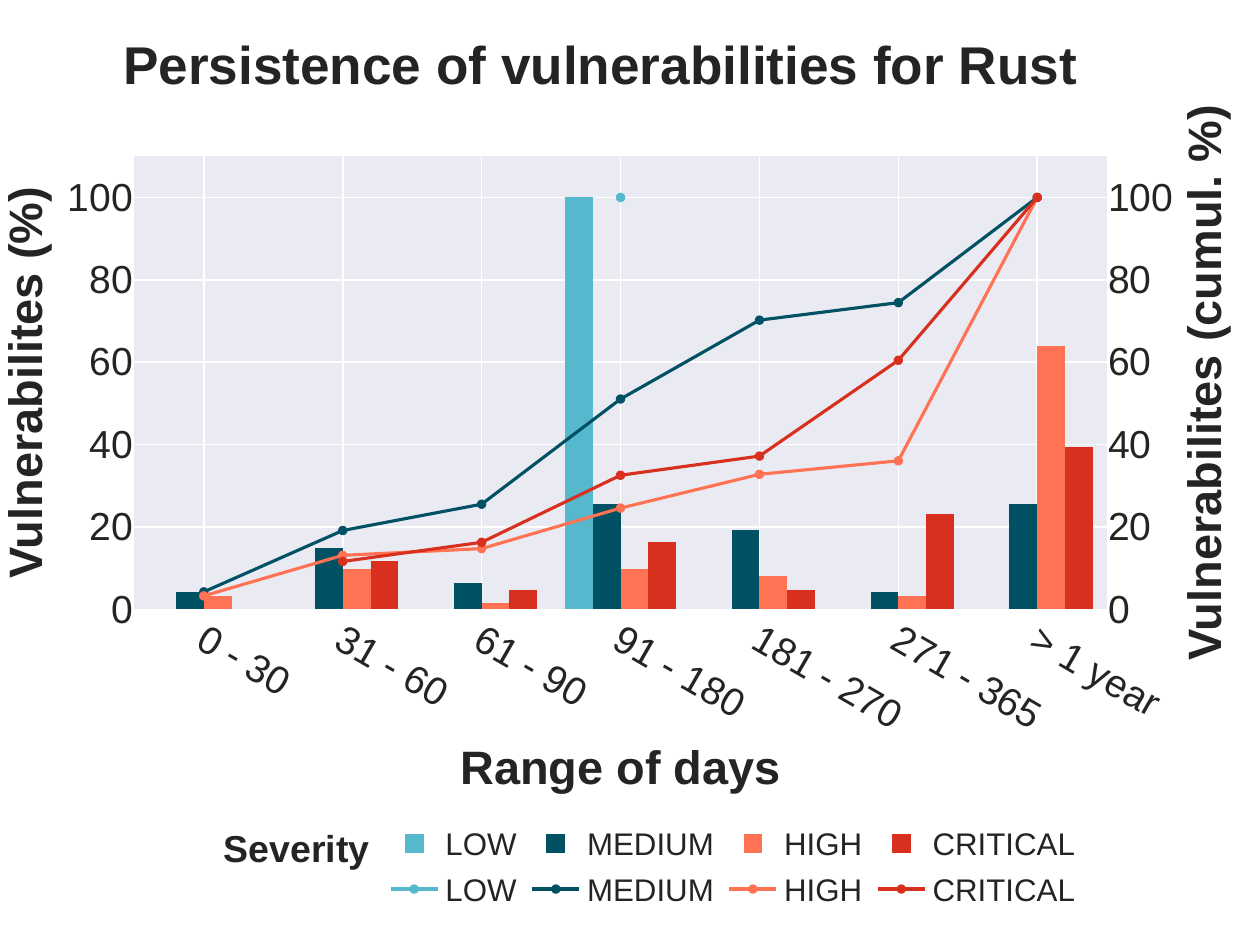}
        \caption{Rust.}
        \label{fig:persistence-rust}
    \end{subfigure}
    \begin{subfigure}[t]{0.24\textwidth}
        \includegraphics[width=\textwidth]{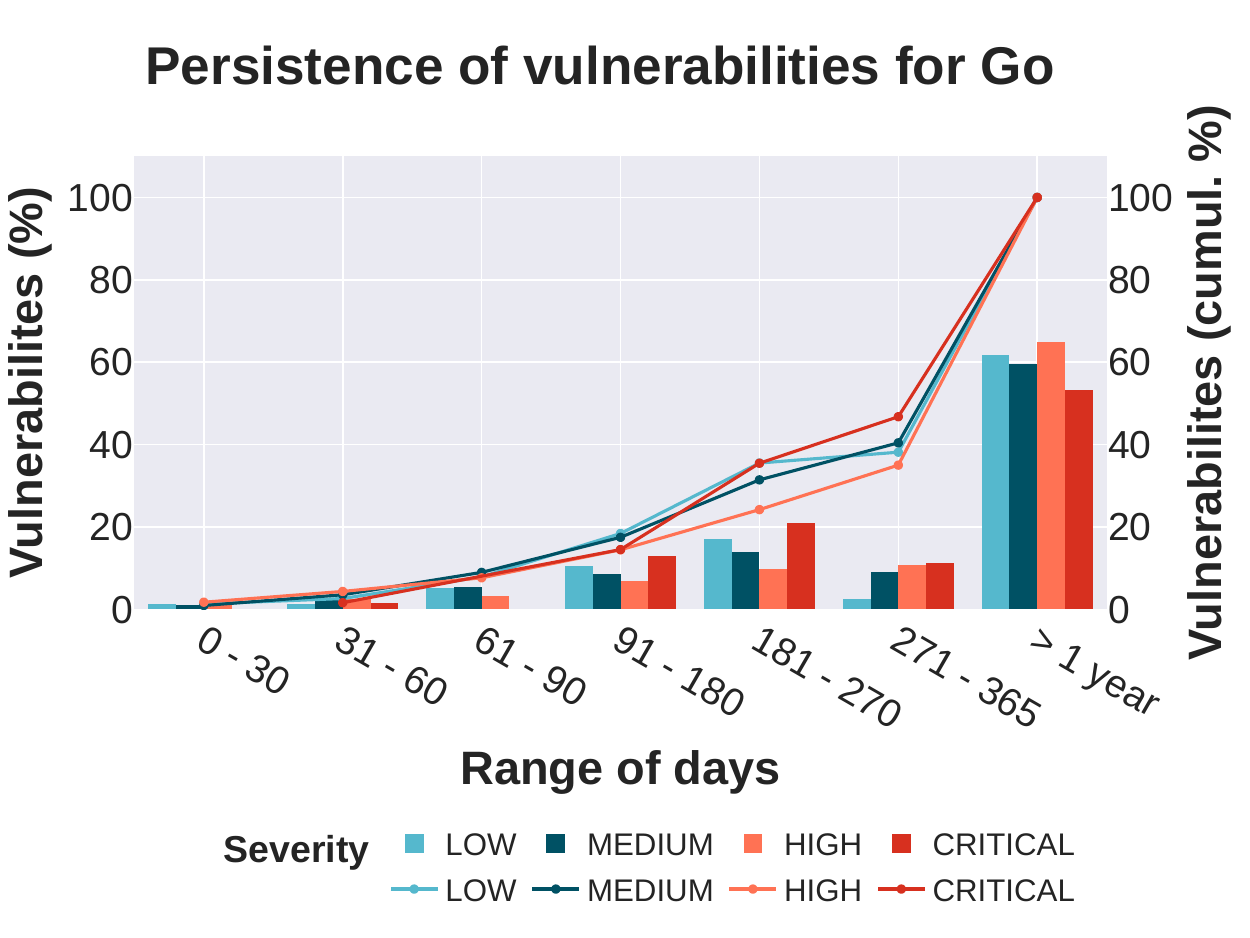}
        \caption{Go.}
        \label{fig:persistence-go}
    \end{subfigure}
    \begin{subfigure}[t]{0.24\textwidth}
        \includegraphics[width=\textwidth]{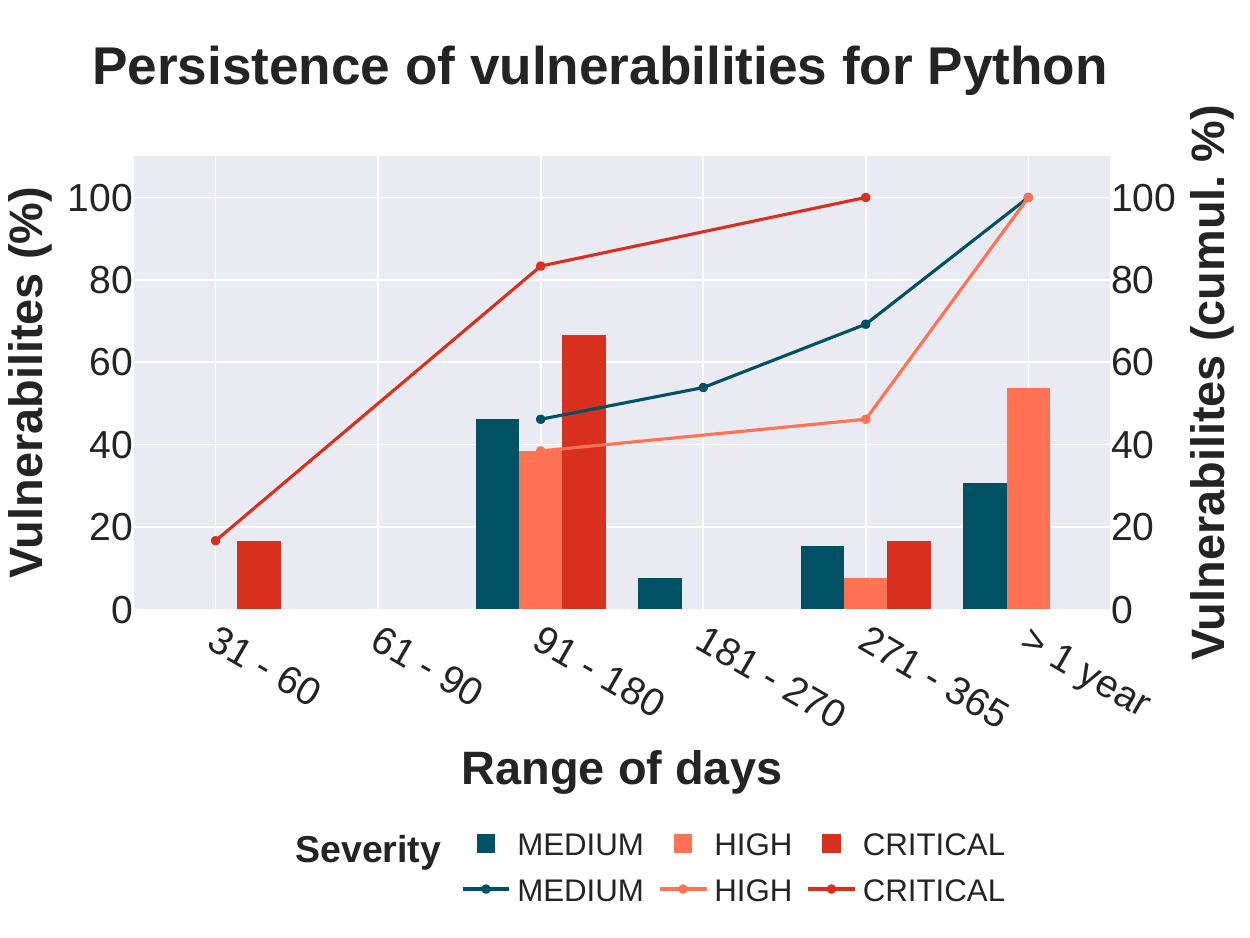}
        \caption{Python.}
        \label{fig:persistence-python}
    \end{subfigure}
    \caption{Cumulative persistence of vulnerabilities.}
    \label{fig:vuln-persistence}
    \vspace{-1.5em}
\end{figure}

In addition to the correlation analysis, we were also interested in how long vulnerabilities persist across repositories and if that reveals a certain trend such that they persist for an extended period. To conduct this analysis, we consider repositories generated from 2013 onward, including the most recent ones. For each repository, we identify the initial release in which a vulnerability was reported and the subsequent release in which it was not reported anymore. By calculating the number of days between these two releases, we determine the persistence in days of a vulnerability. This technique is applied uniformly to all repositories, regardless of programming language, considering only disclosed and patched vulnerabilities. The average duration of vulnerability persistence, categorized by severity, for each programming language, is presented in Figure~\ref{fig:vuln-persistence}. Note that the lines are cumulative while the bars are the percent.
This figure depicts how long a vulnerability persists in a software project until it gets fixed. We partitioned the data by severity to see if high risk vulnerabilities are fixed sooner or stay equally long in the project which seems to be the case.
The severity is furthermore determined based on the CVSS values derived from the NVD data.



From the obtained results, we can conclude that developers in the open-source ecosystem generally may not actively upgrade components to the most recent versions, resulting in vulnerabilities persisting for extended periods. Therefore, caution should be exercised when selecting open-source projects, and known vulnerabilities should be checked. If a vulnerability is found, the CVSS and EPSS scores can be examined to determine the likelihood of exploitation and make decisions accordingly.

\section{Related Work} \label{sec:related}

There has been numerous works \cite{Prana2021,Pashchenko2022,Pashchenko2018,Kula2018,Cadariu2015,Ponta2020} that aim to better understand and study vulnerable components within the open-source ecosystem. 

An empirical analysis was conducted by Prana et al.~\cite{Prana2021} on $450$ GitHub projects to identify common vulnerabilities, vulnerable components, and the persistence of vulnerabilities using the Veracode SCA tool. The study focused on Java, Python, and Ruby repositories with at least five commits over the course of one year, from 2017 to 2018. 
Common vulnerability types and vulnerable libraries that impact numerous projects were identified. Additionally, it was discovered that the number of vulnerabilities associated with open-source dependencies tends to be higher in Java and Ruby projects. Other findings indicate that a significant percentage of vulnerable dependency issues persist, and among those that are resolved, the average time taken is approximately $4-5$ months. Besides that the number of direct and transitive dependencies strongly correlates with vulnerability counts. This emphasizes the importance for library users to carefully manage the number of dependencies in their projects and ensure timely updates.

Ponta et al.~\cite{Ponta2020} introduces a new technique for identifying and mitigating vulnerabilities in open-source software that focuses on the code itself rather than metadata. Their approach combines static and dynamic analyses to determine the accessibility of vulnerable parts of libraries within the context of a specific application. The solution also helps developers select compatible versions of non-vulnerable libraries to minimize potential incompatibilities. Eclipse Steady, the open-source implementation of their approach, is recommended as the tool of choice for scanning Java software products at SAP. An empirical study comparing Eclipse Steady with OWASP Dependency Check found that all findings from Steady were true positives, while $88.8\%$ of the findings from OWASP were false positives for vulnerabilities covered by the code-centric approach.

Besides that, other works such as Cadariu et al.~\cite{Cadariu2015} conducted research on the prevalence of known vulnerable dependencies in $75$ proprietary Java projects built with Maven. Their findings revealed that $54$ of the projects utilized at least one library with known security vulnerabilities. Pashchenko et al.~\cite{Pashchenko2018, Pashchenko2022} proposed the Vuln4Real methodology which offers a reliable approach to measuring of vulnerable dependencies in OSS libraries. 
To prove its effectiveness, the researchers applied Vuln4Real to the top $500$ OSS Maven-based libraries from SAP. They developed a tool that uses Apache Maven to extract library dependencies and implement Vuln4Real post-processing steps. The results show that Vuln4Real significantly impacts the ecosystem and individual library developers. Previous research also showed that about $20\%$ of affected dependencies were not deployed, and $81\%$ of vulnerable dependencies could be fixed by simply updating the library version. 

Several works exists that is focused only on one particular language and its dependency structure. For examples, Zimmermann et al.~\cite{Zimmermann2019} investigate the inter package dependencies and their vulnerabilities of npm packages. While the paper reveals that only a small number of heavily used and unmaintained npm packages can lead to large and catastrophic impacts, the study is limited to the JavaScript language and its npm ecosystem. 

Likewise, the Mir et al.~\cite{Mir2023} investigate vulnerabilities, its reach-ability and call graph structure on the maven ecosystem limiting the scope to the JAVA programming language. While the focus of this work was to determine the percentage packages which are vulnerable, our study complements these insights with a historical analysis across several languages.

With regards to a historical analysis, Zheng et al.~\cite{Zheng2023} performed a study on the Rust ecosystem/crates and analyzed 300 vulnerable code repositories including its seven years history. While the study is quite similar to ours such as characterizing the popularity, categorization, and vulnerability density, the study is limited to the Rust programming language. 

In a similar way, the study of Kula et al.~\cite{Kula2018} is also limited to Java while the research questions are similar to ours such as gaining a better understanding of developers' behaviour such as updating dependencies if security advisories have been published.

There exists also a large body of works that is focused on slightly different aspects such as tracking patches rather than vulnerabilities in software dependencies. Hence, the Xu et al.~\cite{Xu2022} are targeting the quality of patches rather than investigating how wide spread software vulnerabilities are due to complex dependency trees and transitivity.

Several existing tools can be utilized for crawling and analysing dependencies. For instance the DEPHEALTH tool as mentioned in the paper of Alfadel et at.~\cite{Alfadel2021} can be used to crawl Python packages. However, the tool is not publicly available, hence not allowing us to use it to conduct similar studies.

In summary, our research uses a larger and more diverse dataset in comparison to previous studies on vulnerable dependency usage. The dataset consists of various software projects with different characteristics, encompassing multiple programming languages such as Java, Python, Rust, Go, Ruby, PHP, and JavaScript, which enhances the generalizability of our findings.

\section{Conclusions} \label{sec:conclusions}
This paper presented VODA (\vodaFullName), a distributed system consisting of a highly parallel and scalable crawler infrastructure for \textit{git} repositories as well as an analytics pipeline to scan and analyze source code repositories for vulnerabilities. Using \voda{}, we were able to analyze more than $50k$ software releases across $1k$ distinct source code repositories to present a comprehensive study that provides valuable insights about vulnerability persistence and several metrics and correlations such as contributors and team sizes and how maintenance and other factors affects vulnerability removal.

As future work, we plan to extend our system to also include C/C++ repositories, which is non-trivial as there exist no standard package/dependency management systems for proper SBOM generation. Instead, it requires inferring dependencies from Makefiles and configuring scripts that do not contain any version information and, hence, are error-prone and quite cumbersome. Also, we want to explore how this knowledge can be used to improve compliance processes and tools.

\section*{Acknowledgments}\label{sec:acknowledgements}

This work was supported in part by Deutsche Forschungsgemeinschaft (DFG, German Research Foundation) as part of the Cluster of Excellence Centre for Tactile Internet with Human-in-the-Loop (CeTI) -- Project ID 390696704, and the CRC/Transregio 248 Foundations of Perspicuous Software Systems (CPEC) -- Project ID 389792660.
The authors also acknowledge the financial support by the Federal Ministry of Education and Research of Germany in the programme of "Souverän. Digital. Vernetzt.". Joint project 6G-life, project identification number: 16KISK001K,
and the European Union Horizon Europe research and innovation programme under grant agreements 101092644 (NEARDATA) and 101092646 (CLOUDSKIN).

\balance
\bibliographystyle{bib/IEEEtran}
\bibliography{bib/ieee}

\end{document}